\documentclass[longauth, a4]{aa} 

\usepackage{graphicx}
\usepackage{natbib}
\bibpunct{(}{)}{,}{a}{}{;}
\usepackage{amssymb}
\usepackage{amsmath}
\usepackage{url, verbatim}
\usepackage{multirow}
\usepackage{txfonts}
\newcommand{\phs}{\phantom{1}}
\newcommand{\phn}{\phantom{--}}

\begin{document}
   \title{Origin of warm and hot gas emission from low-mass protostars: \textit{Herschel}-HIFI observations of CO \textit{J}=16--15\thanks{\textit{Herschel} is an ESA space observatory with science instruments provided by European-led Principal Investigator consortia and with important participation from NASA.}}
\subtitle{I. Line profiles, physical conditions, and H$_2$O abundance}
\titlerunning{I. Line profiles, physical conditions, and H$_2$O abundance}


\author{L.E.~Kristensen\inst{1,2}
\and E.F.~van~Dishoeck\inst{2,3}
\and J.C.~Mottram\inst{4}
\and A.~Karska\inst{5}
\and U.A.~Y{\i}ld{\i}z\inst{6}
\and E.A.~Bergin\inst{7}
\and P.~Bjerkeli\inst{1}
\and S.~Cabrit\inst{8}
\and S.~Doty\inst{9}
\and N.J.~Evans II\inst{10}
\and A.~Gusdorf\inst{11}
\and D.~Harsono\inst{12}
\and G.J.~Herczeg\inst{13}
\and D.~Johnstone\inst{14,15}
\and J.K.~J{\o}rgensen\inst{1}
\and T.A.~van~Kempen\inst{16,2}
\and J.-E.~Lee\inst{17,18}
\and S.~Maret\inst{19,20}
\and M.~Tafalla\inst{21}
\and R.~Visser\inst{22}
\and S.F.~Wampfler\inst{23}
}

\institute{
Centre for Star and Planet Formation, Niels Bohr Institute and Natural History Museum of Denmark, University of Copenhagen, {\O}ster Voldgade 5-7, DK-1350 Copenhagen K, Denmark, \email{lars.kristensen@nbi.ku.dk}  \and
Leiden Observatory, Leiden University, PO Box 9513, 2300 RA Leiden, the Netherlands \and
Max Planck Institut f\"{u}r Extraterrestrische Physik, Giessenbachstrasse 1, 85748 Garching, Germany \and
Max Planck Institute for Astronomy, K{\"o}nigstuhl 17, 69117 Heidelberg, Germany \and
Centre for Astronomy, Nicolaus Copernicus University, Faculty of Physics, Astronomy and Informatics, Grudziadzka 5, PL-87100 Torun, Poland \and
Jet Propulsion Laboratory, 4800 Oak Groave Drive, Pasadena, CA 91109, USA \and
Department of Astronomy, The University of Michigan, 500 Church Street, Ann Arbor, MI 48109-1042, USA \and
LERMA, UMR 8112 du CNRS, Observatoire de Paris, 61 Av. de l'Observatoire, 75014 Paris, France \and
Department of Physics and Astronomy, Denison University, Granville, OH, 43023, USA \and
Department of Astronomy, The University of Texas at Austin, Austin, TX 78712, USA \and
LERMA, UMR 8112 du CNRS, Observatoire de Paris, {\'E}cole Normale Sup{\'e}rieure, 61 Av. de l'Observatoire, 75014, Paris, France \and
Heidelberg University, Center for Astronomy, Institute for Theoretical Astrophysics, Albert-Ueberle-Strasse 2, 69120 Heidelberg, Germany \and
Kavli Institute for Astronomy and Astrophysics, Peking University, Yi He Yuan Lu 5, Haidian Qu, 100871 Beijing, China \and
National Research Council Canada, Herzberg Institute of Astrophysics, 5071 West Saanich Road, Victoria, BC V9E 2E7, Canada
\and
Department of Physics and Astronomy, University of Victoria, Victoria, BC V8P 1A1, Canada \and
SRON Netherlands Institute for Space Research, Location Utrecht, Sorbonnelaan 2, 3584 CA, Utrecht \and
Department of Astronomy \& Space Science, Kyung Hee University, Gyeonggi 446-701 \and
Korea School of Space Research, Kyung Hee University, Yongin-shi, Kyungki-do 449-701, Korea \and
Univ. Grenoble Alpes, IPAG, 38000 Grenoble, France \and
CNRS, IPAG, 38000 Grenoble, France \and
Observatorio Astron\'{o}mico Nacional (IGN), Calle Alfonso XII,3. 28014, Madrid, Spain \and
 European Southern Observatory, Karl-Schwarzschild-Strasse 2, 85748 Garching, Germany \and
Center for Space and Habitability, University of Bern, Sidlerstrasse 5, CH-3012 Bern, Switzerland
}

\date{Draft: \today}

\abstract
{Through spectrally unresolved observations of high-$J$ CO transitions, \textit{Herschel} Photodetector Array Camera and Spectrometer (PACS) has revealed large reservoirs of warm (300 K) and hot (700~K) molecular gas around low-mass protostars. The excitation and physical origin of this gas is still not understood.}
{We aim to shed light on the excitation and origin of the CO ladder observed toward protostars, and on the water abundance in different physical components within protostellar systems using spectrally resolved \textit{Herschel} Heterodyne Instrument for the Far Infrared (HIFI) data.}
{Observations are presented of the highly excited CO line $J$ = 16--15 ($E_{\rm up}$/$k_{\rm B}$ = 750 K) with \textit{Herschel}-HIFI toward a sample of 24 low-mass protostellar objects. The sources were selected from the \textit{Herschel} ``Water in Star-forming regions with \textit{Herschel}'' (WISH) and ``Dust, Ice, and Gas in Time'' (DIGIT) key programs.}
{The spectrally resolved line profiles typically show two distinct velocity components: a broad Gaussian component with an average $FWHM$ of 20 km\,s$^{-1}$ containing the bulk of the flux, and a narrower Gaussian component with a $FWHM$ of 5 km\,s$^{-1}$ that is often offset from the source velocity. Some sources show other velocity components such as extremely-high-velocity features or ``bullets''. All these velocity components were first detected in H$_2$O line profiles. The average rotational temperature over the entire profile, as measured from comparison between CO $J$=16--15 and 10--9 emission, is $\sim$ 300 K. A radiative-transfer analysis shows that the average H$_2$O/CO column-density ratio is $\sim$0.02, suggesting a total H$_2$O abundance of $\sim$2$\times$10$^{-6}$, independent of velocity.}
{Two distinct velocity profiles observed in the HIFI line profiles suggest that the high-$J$ CO ladder observed with PACS consists of two excitation components. The warm PACS component (300 K) is associated with the broad HIFI component, and the hot PACS component (700 K) is associated with the offset HIFI component. The former originates in either outflow cavity shocks or the disk wind, and the latter in irradiated shocks. The low water abundance can be explained by photodissociation. The ubiquity of the warm and hot CO components suggests that fundamental mechanisms govern the excitation of these components; we hypothesize that the warm component arises when H$_2$ stops being the dominant coolant. In this scenario, the hot component arises in cooling molecular H$_2$-poor gas just prior to the onset of H$_2$ formation. High spectral resolution observations of highly excited CO transitions uniquely shed light on the origin of warm and hot gas in low-mass protostellar objects.}


\maketitle

\section{Introduction}

\begin{table*}[!t]
\caption{Characteristics of distinct kinematical and excitation components seen in high-$J$ CO emission. \label{tab:component}}
\small
\begin{center}
\begin{tabular}{l l l l c} \hline \hline
Instrument & Component\tablefootmark{a} & Characteristics\tablefootmark{b} & Possible origin\tablefootmark{c} & References \\ \hline
HIFI & Broad & $FWHM$ $\gtrsim$ 10--15~km\,s$^{-1}$ & Outflow cavity shocks & 1, 2, 3 \\
& & $\varv$ $\sim$ $\varv_{\rm source}$ & MHD disk wind & 4, 5 \\
& Offset & $FWHM$ $\sim$ 5--40~km\,s$^{-1}$ & Spot shock (irradiated) & 1, 2, 3, 6, 7 \\ 
& & $|\varv - \varv_{\rm source}|$ $>$ 1~km\,s$^{-1}$ & EHV bullets & 2, 3 \\
PACS & Warm & $T_{\rm rot}$ $\sim$ 300K & UV-heated outflow cavity walls & 8, 9, 10 \\
& & 14 $<$ $J_{\rm up}$ $<$ 24 & Warm shock & 11, 12, 13 \\
& Hot & $T_{\rm rot}$ $\sim$ 600--800K & C-type shock (irradiated) & 8, 9, 14 \\
& & 24 $<$ $J_{\rm up}$ $<$ 39 & Single shock connecting with warm component & 15, 16, 17 \\ \hline
\end{tabular}
\tablefoot{
\tablefoottext{a}{These components will henceforth be referred to as the broad, offset, warm, and hot components.}
\tablefoottext{b}{Defining observed characteristics; defining observed characteristics are from \citet{mottram14} for the HIFI components, and \citet{karska13} for the PACS components.}
\tablefoottext{c}{Physical origin as proposed in the cited references.}
}
\tablebib{(1)~\citet{kristensen10}; (2)~\citet{kristensen12}; (3)~\citet{mottram14}; (4)~\citet{panoglou12}; (5)~\citet{yvart16}; (6)~\citet{kristensen13}; (7)~\citet{benz16}; (8)~\citet{vankempen10}; (9)~\citet{visser12}; (10)~\citet{lee14}; (11)~\citet{herczeg12}; (12)~\citet{goicoechea12}; (13)~\citet{karska13}; (14)~\citet{karska14}; (15)~\citet{neufeld12}; (16)~\citet{manoj13}; (17)~\citet{green13}. }
\end{center}
\end{table*}

It was a surprise when the \textit{Herschel} Photodetector Array Camera and Spectrometer (PACS) revealed that low- and high-mass embedded protostars host a reservoir of warm and hot material with temperatures $\gtrsim$ 300 K, as seen in observations of highly excited CO emission \citep[$J_{\rm up}$ $>$ 14;][Karska et al. in prep.]{karska13, manoj13, green13, karska14, matuszak15, green16, manoj16}. The origin of this warm/hot gas is still not known partly because of uncertainty over the excitation conditions of CO (density and temperature) and because the physical processes at play on small scales ($<$ 500 AU) in low-mass protostars are still not completely understood. Few direct observations of this warm material exist because the ubiquitous tracer, H$_2$, is not directly observable toward the deeply embedded low-mass protostars, due to the high extinction \citep[e.g.,][]{maret09} often exceeding an $A_{\rm V}$ of 100--1000. 

\textit{Herschel}-PACS observations of the CO ladder of low-mass protostars ranging from $J$ = 14--13 to 49--48 ($E_{\rm up}$/$k_{\rm B}$~=~500--6500~K) typically show two rotational temperature components: a warm component with a rotational temperature, $T_{\rm rot}$, of $\sim$ 300 K and a hot component with $T_{\rm rot}$~$\sim$~600--800~K \citep{manoj13, karska13, green13}. For sources with particularly bright CO emission, a third very hot component is sometimes seen with $T_{\rm rot} > 1000$ K \citep{manoj13}, although not always \citep{herczeg12, goicoechea12}. Clearly any successful interpretation must account for both the excitation conditions, and the universality of these components. Determining the physical origin of the CO excitation is an obvious goal, and several groups have undertaken different approaches. These approaches fall into several categories: \textit{(i)} the entire CO ladder is reproduced by a single set of excitation conditions (H$_2$ density, CO column density, and temperature), where the excitation is subthermal \citep{neufeld12, manoj13}; \textit{(ii)} the two temperature components are interpreted as two physically distinct components with different excitation conditions \citep{karska13, green13}; \textit{(iii)} the entire CO ladder is reproduced by a range of temperatures, where the column density of each temperature layer follows a power-law \citep{neufeld12}; or \textit{(iv)} the CO ladder is modeled with a detailed physical model of the entire system \citep{vankempen10, visser12, lee14}. The number of possible solutions demonstrates that the integrated intensities of the CO ladder alone do not provide enough discriminating power to identify the origin of the warm and hot gas. 

subthermal
A complementary warm gas tracer in protostellar systems is H$_2$O \citep[e.g.,][and references therein]{vandishoeck14}. \citet{kristensen13} and \citet{mottram14} recently discussed the excitation and origin of H$_2$O emission, based on velocity-resolved H$_2$O profiles. The H$_2$O line profiles predominantly consist of distinct spectrally resolved components: a broad, $FWHM$ $>$ 20 km\,s$^{-1}$ component centered on the source velocity, and one or more narrower components, $FWHM$ $\sim$ 5--10 km\,s$^{-1}$, offset from the source velocity by $>$ 5 km\,s$^{-1}$ (see Table \ref{tab:component} for component definitions used in this paper).

\citet{kristensen13} and \citet{mottram14} proposed a scenario for interpreting the H$_2$O profile components in which the protostellar wind is neutral and atomic \citep{lizano88, giovanardi92, choi93, lizano95}, and it drives the large-scale entrained, cold ($\lesssim$ 100 K) outflow gas seen in low-$J$ CO ($J$ $\le$ 6--5) emission. Where the wind is currently entraining envelope material, the conditions are conducive to efficient water formation in the gas-phase \citep{bergin98}, as the temperature exceeds 300 K. This shear layer or mixing layer carries the same characteristics as found in shock waves, i.e., the gas is heated collisionally, the density is high, and the gas is accelerated, giving rise to the broad outflow-like component. For these reasons, this layer was named ``outflow cavity shocks'' by \citet{mottram14}. Where the wide-angle low-velocity wind directly impacts the envelope or cavity walls, discrete shocks appear as velocity-offset components in H$_2$O and high-$J$ CO line profiles, typically blue-shifted by $\sim$ 5--10 km\,s$^{-1}$. These are labeled spot shocks \citep{mottram14} and appear with both distinct kinematical and chemical signatures \citep[see above,][]{kristensen13, kristensen16}. In this picture, the warm (300 K) and hot (700 K) CO gas observed with PACS have two different physical origins, the outflow cavity shocks and spot shocks, respectively. 

An alternative to this scenario is that the water emission directly traces the disk wind \citep{panoglou12, yvart16}. In this scenario, the dusty wind is gently accelerated over 10-AU scales, and this gentle acceleration does not lead to collisional dissociation of molecules in the wind. Furthermore, the wind is dense enough to shield molecules within the wind from the dissociating UV radiation from the accreting protostar during the most embedded stages of star formation, the so-called Class 0 stage \citep{andre90}. At later evolutionary stages, the density in the wind decreases and UV photons from the accreting protostar start to dissociate more and more wind material, thus changing the chemical make-up of this material \citep{nisini15}. In this scenario, the bulk of observed emission is associated with the wind. 

Water suffers from the drawback that its abundance is difficult to measure because the denominator of the $N$(H$_2$O)/$N$(H$_2$) ratio is uncertain; H$_2$ is not detected toward the protostellar position. Instead, CO is often used as a proxy for H$_2$. Previous attempts at measuring the abundance with respect to low-$J$ CO have provided values in the range of 10$^{-8}$--10$^{-4}$ with respect to H$_2$, when assuming a CO/H$_2$ abundance of 10$^{-4}$ \citep[e.g.,][]{franklin08, kristensen12}. Part of the reason for this large spread is that low-$J$ CO and H$_2$O do not trace the same gas, as is evident when their spatial distributions are mapped out \citep{nisini10, tafalla13}. Instead, CO 16--15 and H$_2$O seem to trace the same gas \citep{santangelo13}, and thus CO 16--15 provides a better reference frame for measuring the H$_2$O abundance.

This paper addresses the physical origin of the warm and hot CO emission observed with PACS through velocity-resolved observations of CO~16--15 toward a sample of 24 low-mass embedded protostars. In particular, the question of whether or not the two rotational temperature components correspond to different physical components, as witnessed by different velocity components, will be addressed. The origin of the very hot ($T_{\rm rot}$ $>$ 1000 K) emission is not addressed here, as its contribution to the CO 16--15 line profile is too small to distinguish observationally. Finally the H$_2$O abundance as a function of velocity will be measured using \mbox{CO~16--15} as a reference frame. The mass, momentum, and energetics traced by \mbox{CO 16--15} emission will be measured and compared to that measured from lower-$J$ transitions in a subsequent paper. 

The paper is organized as follows. The \textit{Herschel} Heterodyne Instrument for the Far Infrared (HIFI) observations are presented in Sect. \ref{sect:obs} along with a presentation of the source sample. The results are presented in Sect. \ref{sect:results} and discussed in Sect. \ref{sect:disc}. Concluding remarks are given in Sect. \ref{sect:conclusion}.

\section{Observations}\label{sect:obs}

\begin{table*}
\caption{Sample properties \label{tab:sample}}
\tiny
\begin{center}
\begin{tabular}{l c c c c c c c c c} \hline \hline
Source & RA & Dec & Dist & $L_{\rm bol}$ & $M_{\rm env}$ & CO 10--9 & H$_2$O 2$_{12}$--1$_{01}$ & Extended CO\tablefootmark{a} & Extended H$_2$O\tablefootmark{a} \\
& (h:m:s) & (\degr:\arcmin:\arcsec) & (pc) & ($L_\odot$) & ($M_\odot$) & HIFI & HIFI & PACS & PACS \\ \hline
L1448-MM	& 03:25:38.9	& $+$30:44:05.4	& 235 & 9.0		& 3.9		& x & x &  \\
N1333-IRAS2A	& 03:28:55.6	& $+$31:14:37.1	& 235 & 35.7	& 5.1		& x & x & x \\
N1333-IRAS4A	& 03:29:10.5	& $+$31:13:30.9	& 235 & 9.1		& 5.6		& x & x & x & x \\
N1333-IRAS4B	& 03:29:12.0	& $+$31:13:08.1	& 235 & 4.4		& 3.0		& x & x & x & x \\
BHR71		& 12:01:36.3	& $-$65:08:53.0	& 200 & 14.8	& 2.7		& x & -- & x & x \\
IRAS15398	& 15:43:01.3	& $-$34:09:15.0	& 150 & 1.6		& 0.5		& x & -- & x & x \\
VLA1623		& 16:26:26.4	& $-$24:24:30.0	& 125 & 2.6		& 0.24\tablefootmark{b} & -- & -- & x & x \\
L483			& 18:17:29.9	& $-$04:39:39.5	& 200 & 10.2	& 4.4 	& x & -- &  \\
Ser-SMM1	& 18:29:49.6	& $+$01:15:20.5	& 415 & 99.0	& 16.1 	& x & x &  & x \\
Ser-SMM4	& 18:29:56.6	& $+$01:13:15.1	& 415 & 6.2		& 2.1 	& x & x & x \\
Ser-SMM3	& 18:29:59.2	& $+$01:14:00.3	& 415 & 16.6		& 3.2 	& x & -- & x & x \\
B335			& 19:37:00.9	& $+$07:34:09.6	& 103 & 3.3		& 1.2 	& x & -- &  \\
L1157		& 20:39:06.3	& $+$68:02:15.8	& 325 & 4.7		& 1.5 	& x & -- &  \\ \hline
L1489		& 04:04:43.0	& $+$26:18:57.0	& 140 & 3.8		& 0.2 	& x & -- &  \\
L1551-IRS5	& 04:31:34.1	& $+$18:08:05.0	& 140 & 22.1	& 2.3 	& x & -- &  \\
TMR1		& 04:39:13.7	& $+$25:53:21.0	& 140 & 3.8		& 0.2 	& x & -- &  \\
HH46		& 08:25:43.9	& $-$51:00:36.0	& 450 & 27.9	& 4.4 	& x & -- & x & x \\
DK Cha		& 12:53:17.2	& $-$77:07:10.6	& 178 & 35.4	& 0.8 	& x & -- &  \\
GSS30-IRS1	& 16:26:21.4	& $-$24:23:04.0	& 125 & 13.9	& 0.1 	& x & -- & x & x \\
Elias29		& 16:27:09.4	& $-$24:37:19.6	& 125 & 14.1	& 0.04 	& x & -- &  \\
Oph-IRS44	& 16:27:28.3	& $-$24:39:33.0	& 125 & 0.5		& 0.11\tablefootmark{b} & -- & -- &  \\
RCrA-IRS5A	& 19:01:48.0	& $-$36:57:21.6	& 120 & 7.1		& 2.0 	& -- & -- & x & x \\
RCrA-IRS7C	& 19:01:55.3	& $-$36:57:17.0	& 120 & 9.1		& \ldots 	& -- & -- & x & x \\
RCrA-IRS7B	& 19:01:56.4	& $-$36:57:28.3	& 120 & 8.4		& 2.2\tablefootmark{c} & -- & -- & x & x \\ \hline
\end{tabular}
\tablefoot{Coordinates, distances, luminosities, and envelope masses are either from \citet{kristensen12}, \citet{karska13} or \citet{green13}. CO 10--9 emission is reported in \citet{yildiz13} and H$_2$O 2$_{12}$--1$_{01}$ in \citet{mottram14}; sources not observed in one of the two transitions are marked with ``---''.
\tablefoottext{a}{Spatially extended or not beyond the central PACS spaxel, based on either CO 14--13 \citep{karska13}, or CO 16--15 emission \citep{green13}, for the case of CO, or the H$_2$O 2$_{12}$--1$_{01}$ transition at 179.5 $\mu$m for H$_2$O.}
\tablefoottext{b}{From \citet{enoch09}.}
\tablefoottext{c}{From \citet{lindberg12}.}}
\end{center}
\end{table*}

\subsection{Source sample}

The sources were chosen from the ``Water in Star-forming regions with \textit{Herschel}'' \citep[WISH,][]{vandishoeck11} and ``Dust, Ice, and Gas in Time'' \citep[DIGIT,][]{green13} samples of nearby embedded low-mass protostars. \textit{Herschel}-PACS measurements of the CO 16--15 intensity exist for all sources \citep{green13, karska13}. Specifically, the 21 brightest sources in CO 16--15 were chosen, based on whether the line profile would be detected and spectrally resolved with HIFI in less than one hour of telescope time. Furthermore the sample was augmented by WISH observations of two sources, Ser SMM1 \citep{kristensen13} and DK Cha, and observations of HH46 from a Guaranteed Time program (GT1\_abenz\_1; PI: A.O. Benz). Thus, the total sample consists of 24 embedded sources: 13 Class 0 and 11 Class I. 

The sample characteristics are provided in Table \ref{tab:sample}. HIFI H$_2$O 1$_{10}$--1$_{01}$ spectra at 557 GHz \citep{kristensen12, green13} exist toward all sources. HIFI observations of CO 10--9 exist toward most sources \citep[19/24,][]{yildiz13}, and of H$_2$O 2$_{12}$--1$_{01}$ at 1670 GHz toward a smaller subset \citep[6/24,][]{mottram14}. This H$_2$O line is the closest in frequency to CO 16--15 and is therefore observed in a similar beam \citep{mottram14}.

\begin{figure*}[!t]
\begin{center}
\includegraphics[width=\textwidth]{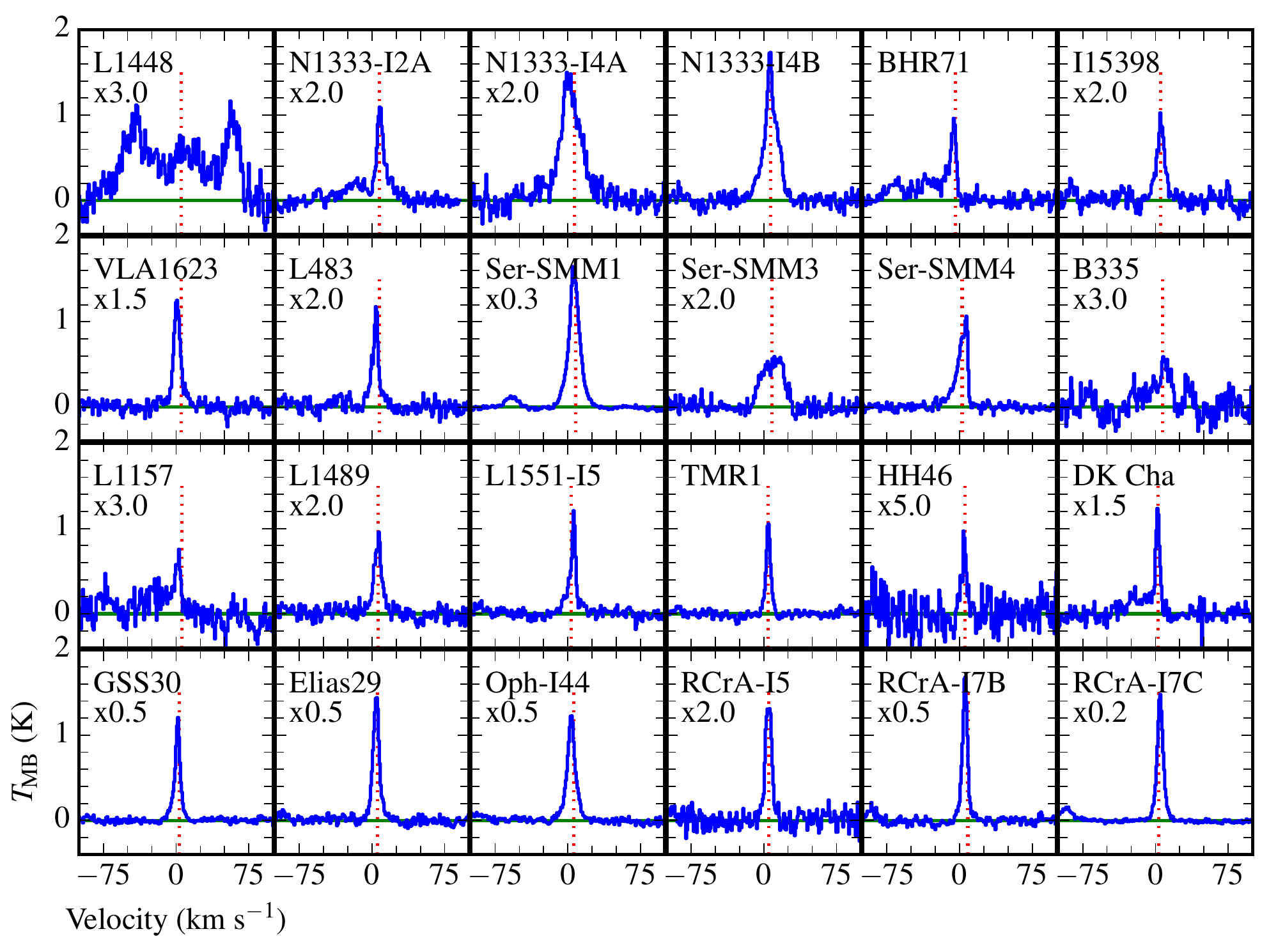}
\caption{CO 16--15 spectra toward all observed sources. The source velocity is marked with a red dashed line in each panel and the baseline is shown in green. Some spectra have been magnified for clarity by a factor shown in each panel. \label{fig:spectra}}
\end{center}
\end{figure*}

\subsection{HIFI data reduction}

The observations consisted of single-pointing dual-beam-switched observations taken with HIFI \citep[the Heterodyne Instrument for the Far Infrared,][]{degraauw10} on \textit{Herschel} \citep{pilbratt10} toward the central position of each of the 24 protostars with the off-position 3$'$ away. The pointing accuracy is 2$''$ (rms). The HIFI spectrometer was tuned to the CO 16--15 line in the upper sideband, and the OH triplet at 1834 GHz in the lower. The OH data will be presented and discussed in a forthcoming publication. The integration time varied per source and was based on the observed line intensity as reported by \citet{green13} and \citet{karska13} (see Table \ref{tab:obsid} in the Appendix for details). The diffraction-limited beam size is 11\farcs5 at the frequency of CO 16--15, corresponding to linear scales of 1400--5200 AU for the distances of the sources observed here. 

Data were reduced using {\sc Hipe} 13.0 \citep{ott10}. Most spectra showed strong signs of standing waves, as expected in HIFI band 7. These waves are not perfectly sinusoidal, and so removing these features requires extra care. The removal was done through the Hipe task \verb+hebCorrection+. 

Furthermore, observations of the hot and cold loads can introduce standing waves with frequencies of 92 and 98 MHz ($\sim$~15 km s$^{-1}$ at the frequency of CO 16--15); these can be eliminated through a modified passband calibration. Additional standing waves with a frequency of $\sim$ 300 MHz ($\sim$ 50 km s$^{-1}$) were removed by using the \verb+fitHifiFringe+ task in Hipe. When running \verb+fitHifiFringe+ the central 1 GHz around the \mbox{CO 16--15} line ($\sim$ 150 km s$^{-1}$) was masked which in most cases covered the entire line. The two sources L1448 and BHR71 are known to harbor extremely high-velocity (EHV) features making the lines as broad as 200 km s$^{-1}$ \citep[full width zero intensity, e.g.,][]{kristensen12} and in these two cases the central 1.5~GHz ($\sim$~250~km\,s$^{-1}$) were masked out of the total bandwidth of 2.5~GHz. The amplitudes of the standing waves are typically 50 mK for the H-polarization spectra and 100 mK for the V-polarization spectra, both on the $T_{\rm A}^*$ scale. These values are comparable to the typical rms level (see below). Some spectra did not show the 300 MHz standing wave, and in these cases no fringes other than the 92/98 MHz ones were removed. Since the periods of the standing waves are similar to the expected line widths (15 and 50 km s$^{-1}$), all spectra and fringe solutions were visually inspected to ensure that the line profiles were not affected. 

After \verb+fitHifiFringe+ had been run on all spectra, they were exported to {\sc Class}\footnote{{\sc Class} is part of the {\sc Gildas} reduction package: \url{http://www.iram.fr/IRAMFR/GILDAS/}} for further reduction and analysis. The reduction consisted of subtracting linear baselines from the spectra and co-adding the H- and V-polarization data after visual inspection. Typically there is a difference in rms level of up to 30\% between the two polarizations. The shape of the profile qualitatively agrees for both the H- and V-polarization data, and so the spectra were averaged. The intensity was brought from the antenna temperature scale, $T_{\rm A}^*$, to the main beam temperature scale, $T_{\rm MB}$, by using a main beam efficiency of 0.60 \citep{roelfsema12}. The calibration uncertainty is measured to $\sim$10\% \citep{roelfsema12}. The spectra were subsequently resampled to a channel width of 0.5~km\,s$^{-1}$. The rms of each spectrum is reported in Table \ref{tab:obsid} in the Appendix. 

\subsection{Complementary data}

The CO 16--15 data presented here are complemented by observations of the H$_2$O 1$_{10}$--1$_{01}$ and 2$_{12}$--1$_{01}$ lines at 557 and 1670~GHz, respectively, also observed with \textit{Herschel}-HIFI \citep{kristensen12, green13, mottram14}. Furthermore, HIFI observations of the CO 10--9 line at 1153 GHz are used \citep{yildiz13}. These data have all been reduced in a similar manner to the CO 16--15 data and the velocity scale was subsequently interpolated to the same scale as the CO 16--15 data for easy comparison.

\section{Results}\label{sect:results}

As expected based on the PACS fluxes, the CO 16--15 line is detected toward every source. Furthermore, the line profiles are spectrally resolved and all spectra are displayed in Fig. \ref{fig:spectra}. Typically the lines are broad ($FWHM$ $\gtrsim$ 15 km s$^{-1}$), and often the profiles consist of multiple dynamical components, similar to what is observed in H$_2$O emission with HIFI \citep{kristensen10, kristensen12, kristensen13, mottram14} and CO 10--9 \citep{yildiz13, sanjosegarcia13}. Comparing the integrated fluxes to those observed with PACS shows that the two fluxes typically agree to better than 20\% which is consistent with the uncertainties on the absolute flux calibration of both instruments. A few sources show larger differences, where the PACS flux is 40--50\% lower than the HIFI flux. This difference may be an timescale of how the PACS emission was extracted and how well-centered the PACS pointing was on the source.

\subsection{Line profiles and profile components}

\begin{figure}[!t]
\begin{center}
\includegraphics[width=\columnwidth]{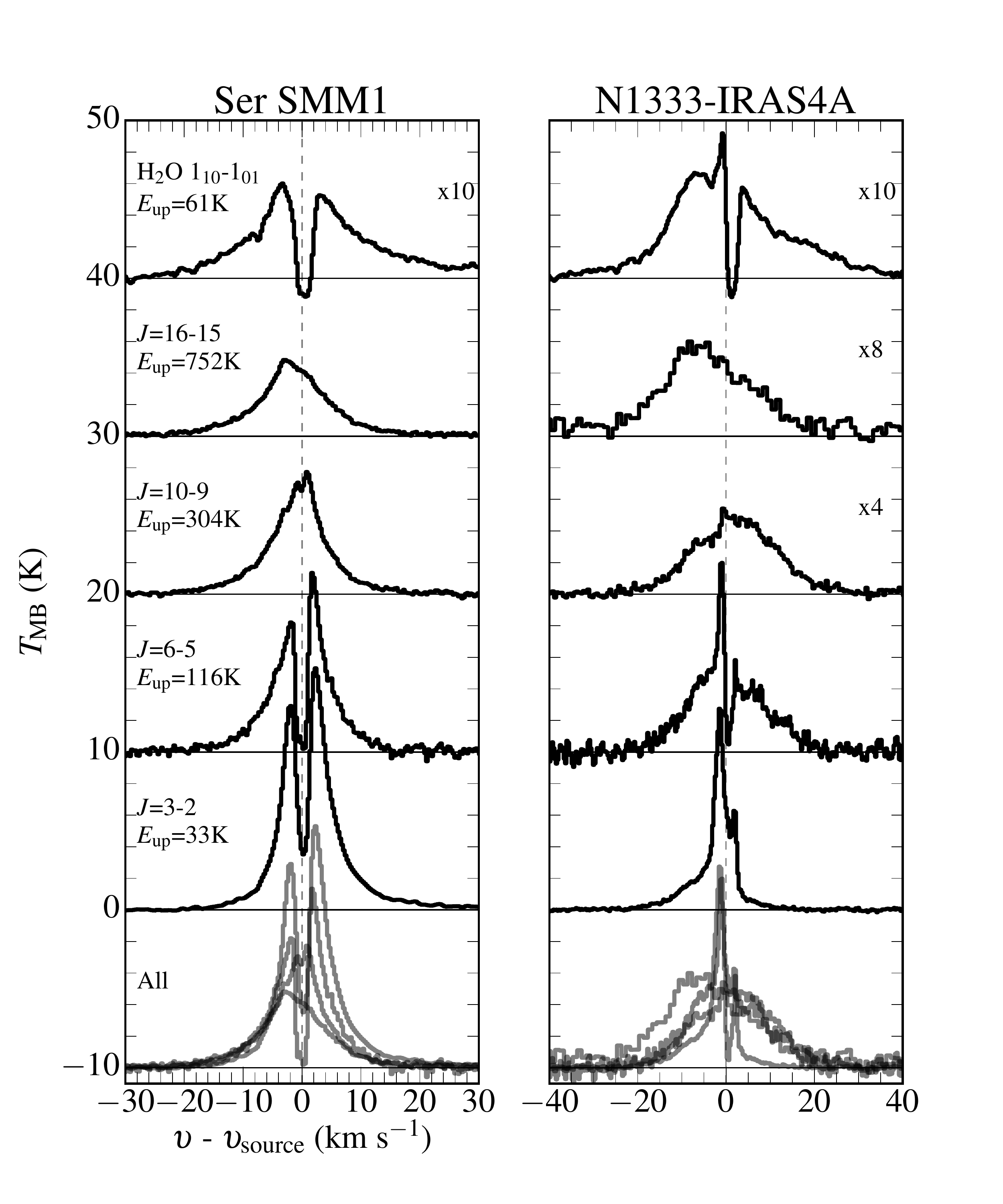}
\caption{CO and H$_2$O spectra toward Ser SMM1 and NGC1333 IRAS4A. No correction has been made for the differences in beam size of the observation. Spectra have been shifted to a velocity of 0 km\,s$^{-1}$. The bottom panel shows all CO profiles overlaid on top of one another. 
\label{fig:smm1_co}}
\end{center}
\end{figure}

\begin{figure}[!t]
\begin{center}
\includegraphics[width=\columnwidth]{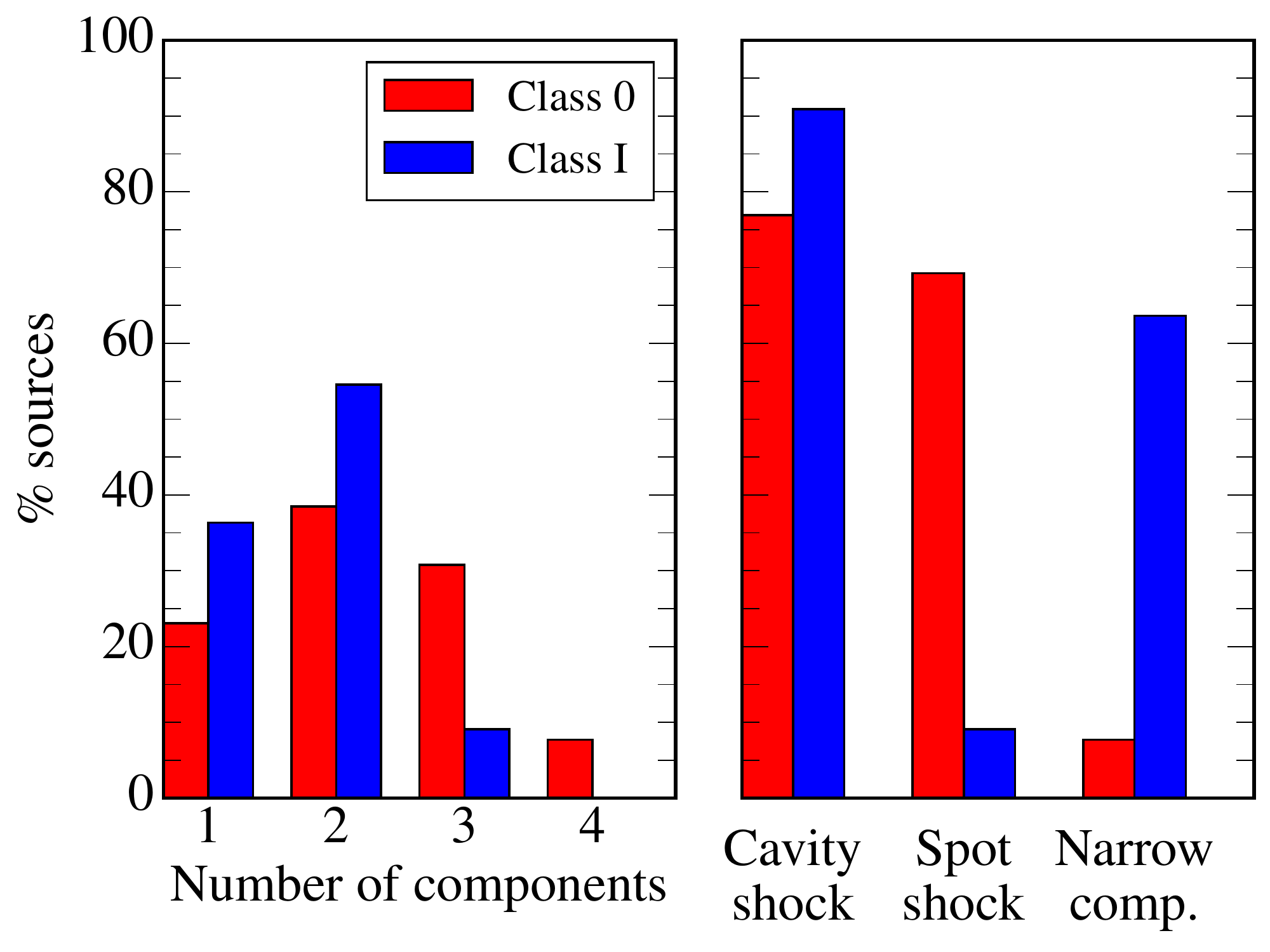}
\caption{Histograms showing the number of Gaussian components in each profile (left) and of profile components (right) toward Class 0 and I sources (red and blue, respectively). \label{fig:comp_histo}}
\end{center}
\end{figure}

The maximum velocity, $\varv_{\rm max}$, traced by the CO line profiles does not change with excitation (see Fig. \ref{fig:smm1_co} for two examples)\footnote{The actual value of $\varv_{\rm max}$ is a function of integration time and $S/N$: when integrating deeper, $\varv_{\rm max}$ always tends to increase \citep{cernicharo89, rudolph92, tafalla10}.}. Instead, CO emission at the lowest velocities decreases with excitation, until only a broad line profile is seen at $J$=16--15. Toward some sources new components start appearing at higher $J$. This is most evident toward NGC1333-IRAS4A and Ser SMM1, where an offset component appears blue-shifted from the source velocity by 10 and 3--4 km\,s$^{-1}$, respectively (Fig. \ref{fig:smm1_co}). These offset components were first detected in H$_2$O \citep{kristensen10, kristensen13}, and only show up in CO transitions with $J_{\rm up}$ $>$ 10 \citep{yildiz13}. 

The line profiles are decomposed using the minimum number of Gaussian functions required for the residual to be less than the rms, following the method of \citet{mottram14}. Gaussian functions reproduce the different components better than, e.g., triangular or Lorentzian functions \citep{sanjosegarcia13}. Two approaches are used: first, a fit is obtained where the Gaussian parameters are all left free; in the second approach, the best-fit parameters ($\varv_0$ and $FWHM$) from \citet{mottram14} are taken as fixed from the decomposition of H$_2$O profiles, and a new fit is obtained. Ser-SMM4 shows a characteristic triangular profile shape, reminiscent of what is seen toward off-source outflow positions; no Gaussian functions are fitted to this profile. All results are listed in Table \ref{tab:gauss} in the Appendix, the spectral decompositions are shown in Fig. \ref{fig:gauss}, also in the Appendix. 

Most sources show multiple components (70\%). Figure \ref{fig:comp_histo} shows the fractional distribution of the number of Gaussian functions needed to reproduce each line profile. This distribution peaks at two components, and then falls off rapidly for higher numbers. Figure \ref{fig:comp_histo} also shows a histogram of the various component detections seen in CO 16--15 and their designations following \citet{mottram14} for H$_2$O. They characterized each component based on its width and its offset from the source velocity and designated them either as ``envelope'' (narrow, centered on $\varv_{\rm source}$), ``cavity shocks'' (broad, centered on $\varv_{\rm source}$), or ``spot shocks'' (offset in velocity). Nearly all sources (85\%) show a broad cavity shock component, irrespective of evolutionary stage, similar to H$_2$O line profiles \citep{mottram14}. Primarily Class 0 sources show offset components associated with spot shocks (70\% versus 10\%) whereas more Class I sources show a narrow component (65\% versus 10\%). 

\begin{figure}[!t]
\begin{center}
\includegraphics[width=\columnwidth]{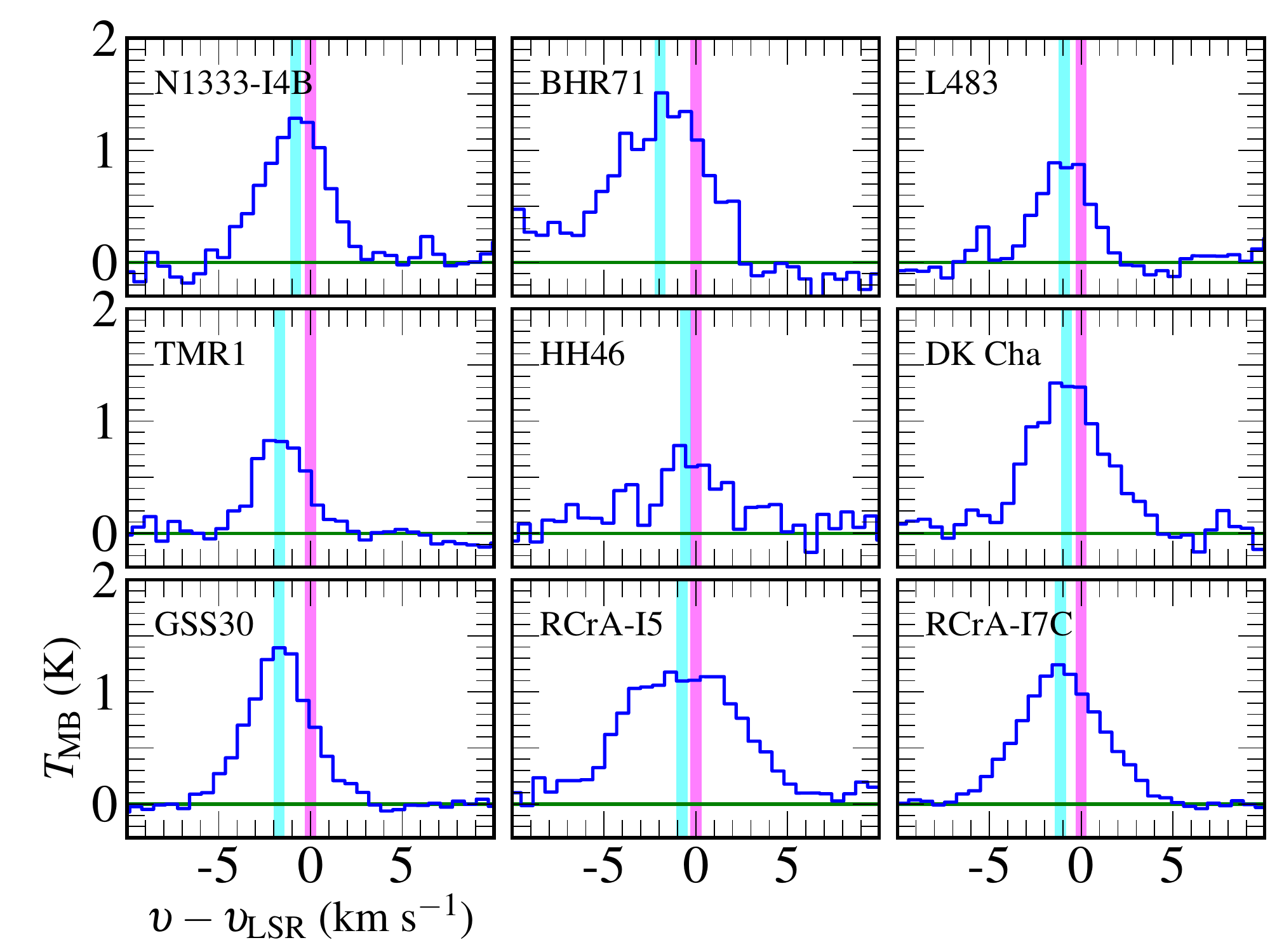}
\caption{Velocity of the narrow component of CO 16--15 with respect to the source velocity where the source velocity is shown in magenta (shifted to 0 km\,s$^{-1}$), the narrow velocity component in cyan, for the nine sources where the component is offset by more than 0.5 km\,s$^{-1}$. Source velocities are measured from low-$J$ C$^{18}$O transitions \citep{yildiz13}. 
\label{fig:narrow}}
\end{center}
\end{figure}

The narrow component is often blue-shifted from the source velocity by 0.5--1.0 km\,s$^{-1}$ (Fig. \ref{fig:narrow}), an offset which is larger than the uncertainty on the velocity calibration \citep[$\ll$0.1 km\,s$^{-1}$,][]{roelfsema12}. The source velocity is measured from low-$J$ C$^{18}$O emission \citep{yildiz13, sanjosegarcia13}. The parts of the envelope traced by higher-$J$ C$^{18}$O 9--8 and $^{13}$CO 10--9 all move at the systemic velocity \citep{yildiz13}. The narrow component is in most cases (10/15) associated with H$_2$O emission, but often blue-shifted by more than 0.5 km\,s$^{-1}$ with respect to the H$_2$O component (7/10 sources); sometimes the shift is larger than 1 km\,s$^{-1}$ (5/10 sources). Of the components not seen in H$_2$O emission, 3/5 components are shifted with respect to the source velocity as measured from low-$J$ C$^{18}$O emission. 

\begin{figure}[!t]
\begin{center}
\includegraphics[width=\columnwidth]{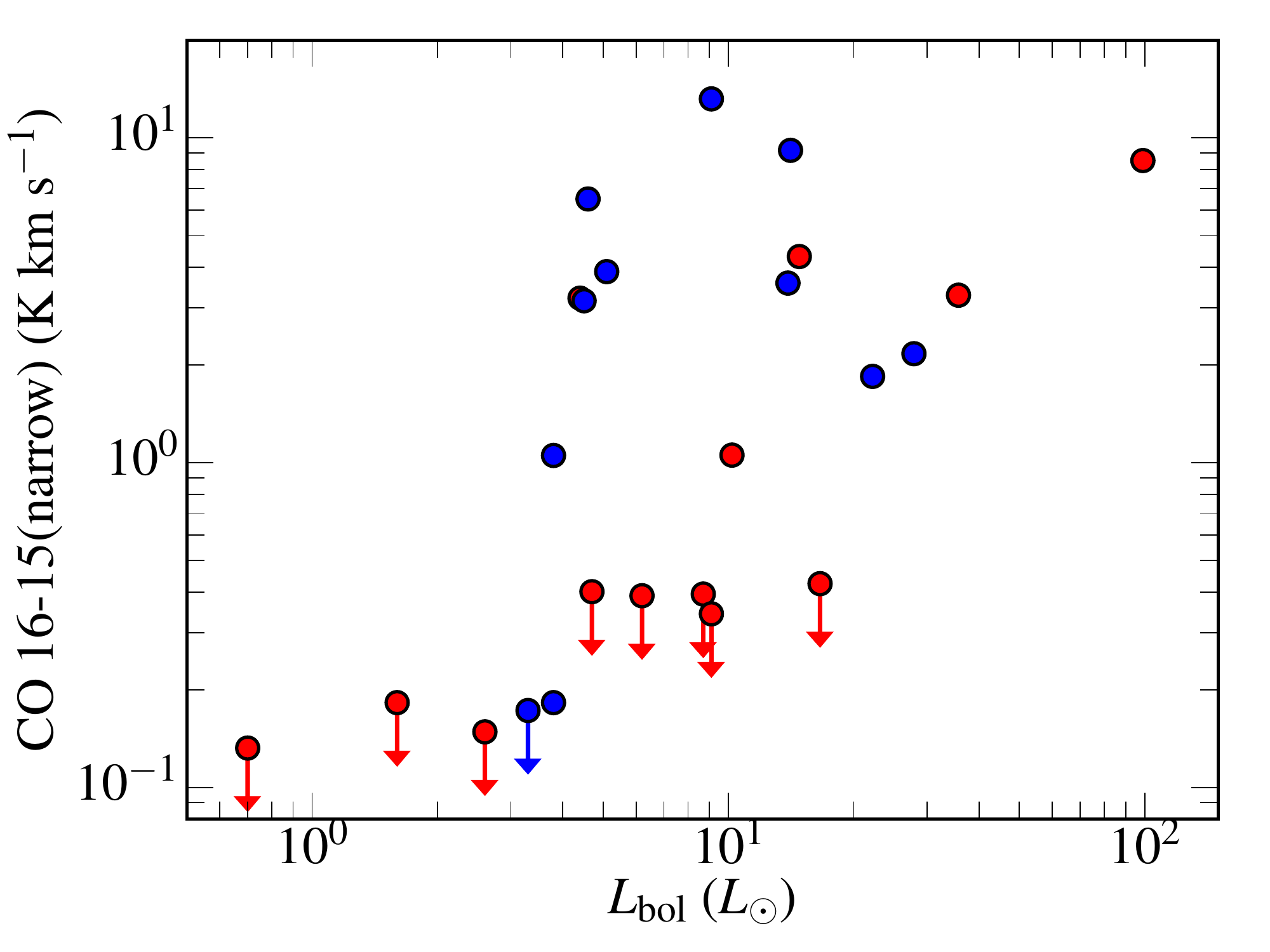}
\caption{\textit{Top:} Narrow emission scaled to a common distance of 200 pc versus $L_{\rm bol}$. Class 0 sources are marked in red, Class I in blue. Upper limits (3$\sigma$) are marked with arrows. Class 0 sources are marked in red, Class I in blue.
\label{fig:outenv}}
\end{center}
\end{figure}

This narrow component is strong enough that if it were present toward all Class 0 sources, it would have been readily detected in the spectra presented here. To illustrate this, figure \ref{fig:outenv} shows the strength of the narrow component as a function of $L_{\rm bol}$ where all upper limits are marked. To calculate the upper limit, the average narrow component line width of 5 km\,s$^{-1}$ is used. The limits are typically an order of magnitude lower than the detected narrow components. Applying Kendall's $\tau$ test with the {\sc Cenken} function\footnote{\url{http://www.practicalstats.com/nada/nadar.html}} from the R statistical package shows that the narrow component strength is correlated with the bolometric luminosity at the 97.1\% or 2.2$\sigma$ level. This test is particularly well suited to evaluate correlations where part of the data set consists of upper limits, i.e., are left-censored. When compared to other envelope parameters (envelope mass, envelope density at 1000 AU, bolometric temperature, and outflow force), there are no correlations, i.e., the significance is $<$ 1$\sigma$ in all cases. Thus the narrow component does not appear to be directly connected to the envelope alone. 

\begin{figure}[!t]
\begin{center}
\includegraphics[width=\columnwidth]{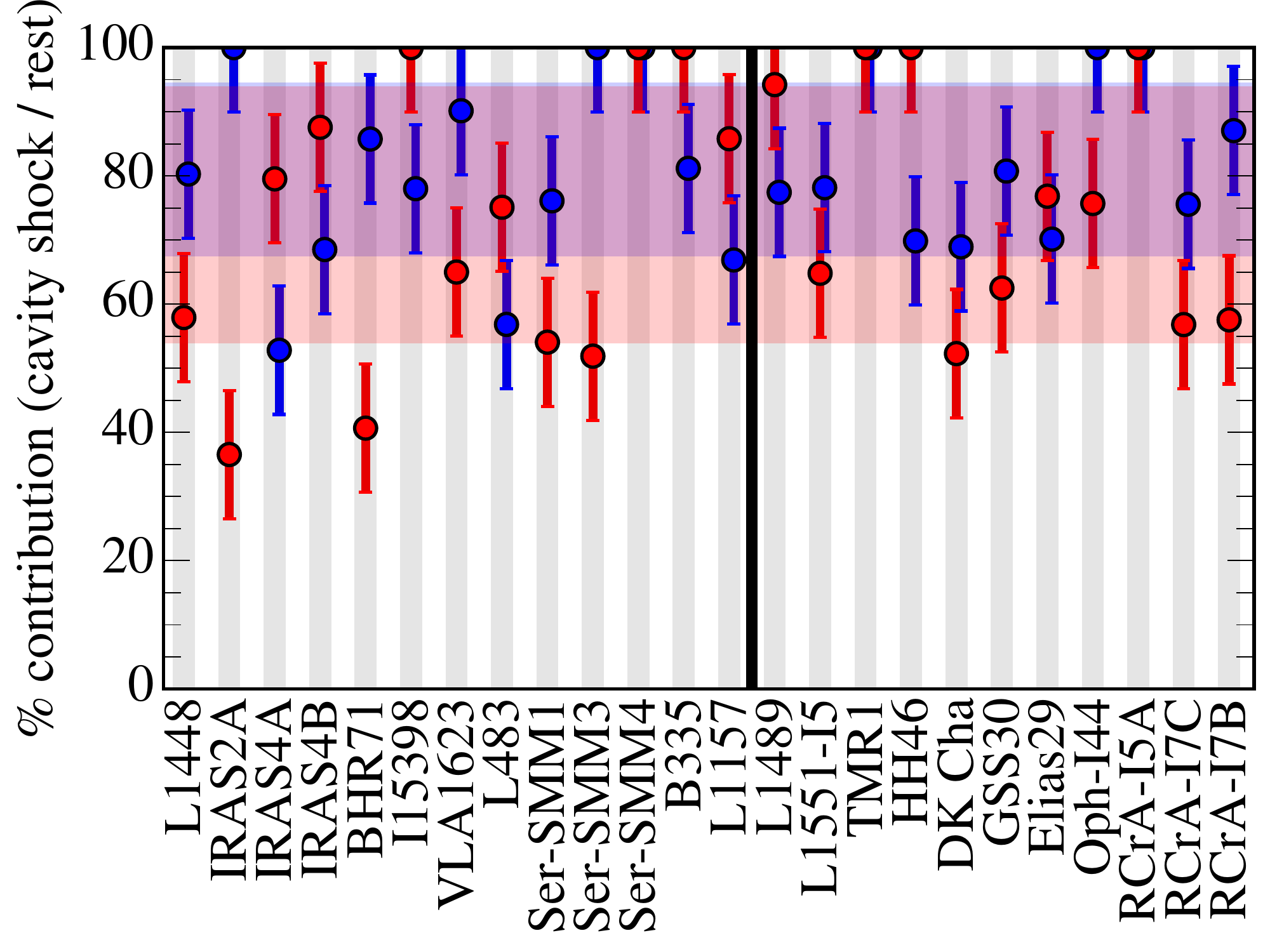}
\caption{Fraction of the contribution to the total integrated CO 16--15 line intensity from the broad cavity shock component as measured from HIFI data (red). The vertical thick black line marks the division between Class 0 and I sources, where the former is to the left, the latter to the right. The blue points show the fraction of CO 16--15 emission expected to originate in the PACS warm 300 K component (Karska et al. in prep.). Both the HIFI and PACS fractions have $\sim$ 10\% uncertainties associated with them. The shaded red and blue regions show the spread of fractions for the HIFI and PACS components, respectively, and where there is an overlap the color is purple. Within the error bars, the fractions are the same, and the broad cavity shock component dominates emission. 
\label{fig:comp_frac}}
\end{center}
\end{figure}

Finally, the fraction of emission in the broad cavity shock component compared to the total integrated intensity is shown in Fig. \ref{fig:comp_frac}. On average, 75$\pm$20\% of the emission is in cavity shocks for all sources. The standard deviation refers to the spread in the percentages; the uncertainty on the decomposition is not included but is smaller. A subset of sources only show emission from the cavity shocks (IRAS15398, Ser-SMM4, B335, TMR1, HH46, and RCrA-IRS5A). An upper limit on any narrower contribution may be estimated assuming its $FWHM$ is 5 km\,s$^{-1}$, the average width of this component in other sources. The 3$\sigma$ limits are up to 20\% for the weakest sources, i.e., consistent with a relative fraction of 75$\pm$20\% in the broad cavity shock component. 

\subsection{Rotational temperature}\label{sect:trot}

The rotational temperature as a function of velocity can be calculated from CO 16--15 and 10--9 spectra, where such data are available (Table \ref{tab:sample}). In the limit where the energy levels are thermally populated, the rotational temperature corresponds to the kinetic temperature; whether this limit applies to these data will be discussed later. To calculate the rotational temperature the spectra are rebinned to a channel size of 3 km\,s$^{-1}$ to increase the $S/N$ ratio. The CO 16--16 / 10--9 line ratio is then calculated in channels where the $S/N$ ratio is $>$ 2 in both spectra and converted into a rotational temperature assuming LTE. The CO 10--9 line may be slightly optically thick at the line center \citep[$\pm$2~km\,s$^{-1}$;][]{yildiz13} toward the brightest sources, and the line centers are ignored in the following analysis. The change in the intensity-weighted average line ratio is less than 10\%, and so not significant compared to the observational uncertainties. 

The CO 10--9 and 16--15 spectra are obtained in different beams, 20\arcsec\ and 11\arcsec, respectively. Maps of CO 10--9 emission show that the emission is typically elongated along the direction of the outflow \citep{nisini15}. Similarly, half of the sources (9/19) where CO 10--9 data are available show extended \mbox{CO 14--13} or 16--15 emission in the PACS footprints \citep{karska13, green13}. If emission from both transitions fill the larger 20\arcsec\ beam, no geometrical correction is required to the line ratio. Alternatively, if the emission follows the outflow a linear scaling with beam size is appropriate \citep{tafalla10}. If the emission is not extended in either of the two beams, the most appropriate geometrical scaling is a point-source scaling. In the following analysis no geometrical scaling will be applied, but the result of possible scalings will be discussed below in Sect. \ref{sect:h2o}. 

\begin{figure}[!t]
\begin{center}
\includegraphics[width=\columnwidth]{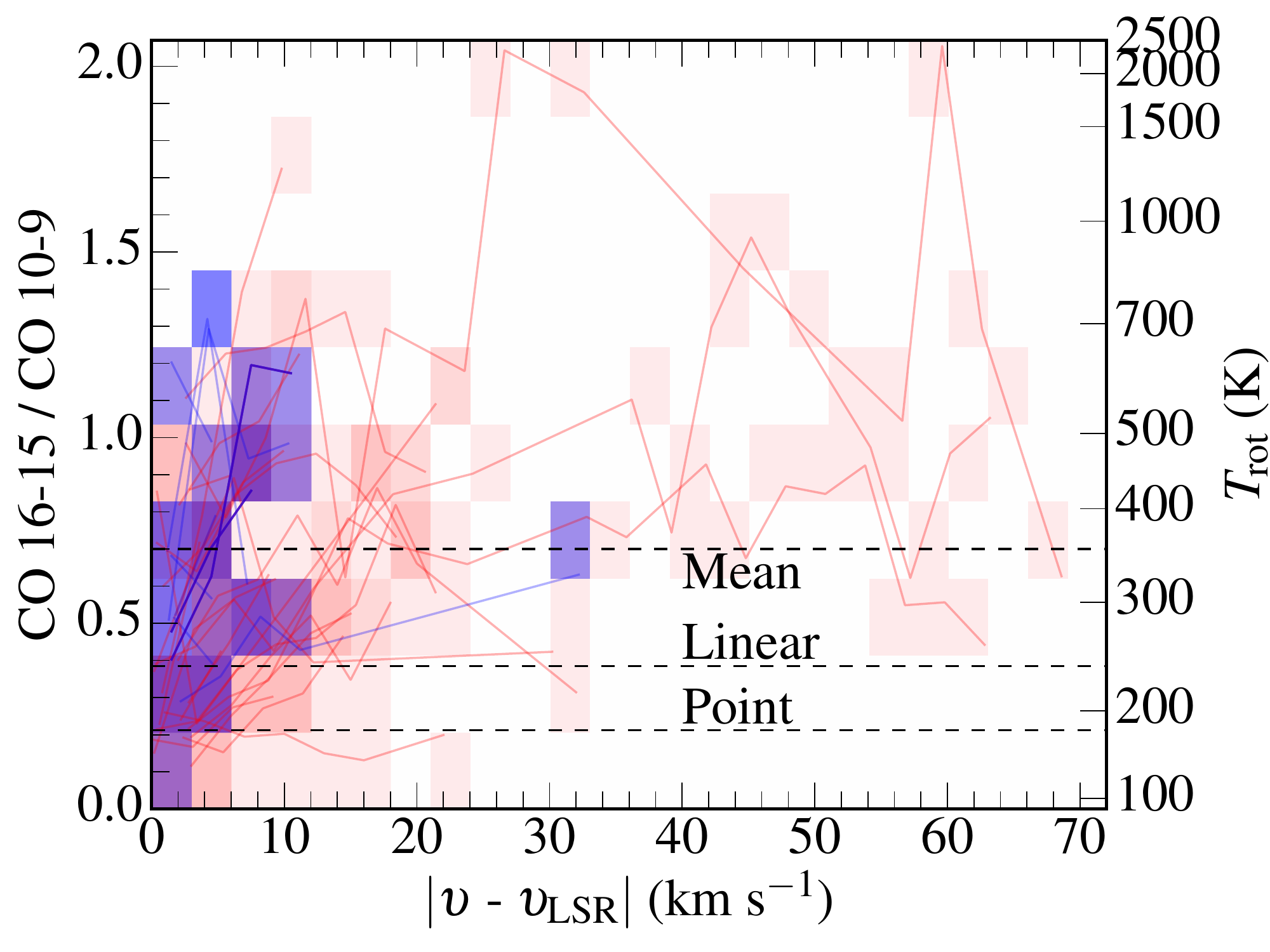}
\caption{CO 16--15 / 10--9 line ratio for all 19 sources for which both data sets are available. The spectra have been rebinned to 3 km\,s$^{-1}$ channels and only data points with $S/N$ $>$ 2 are included. The brightness of the color indicates the number of observations passing through a given point---brighter represents a higher number. Line ratios around $\pm$ 2 km\,s$^{-1}$ are not included. Red is for Class 0 line wings, and blue for Class I. The mean ratio is displayed with a dashed line; if a linear scaling to correct for different beam sizes is applied the mean value shifts down to the dashed line marked ``linear'', and similarly for a point-source scaling. The rotational temperature corresponding to a particular line ratio is shown on the second axis. \label{fig:trot}}
\end{center}
\end{figure}

Figure \ref{fig:trot} shows the average line ratio and corresponding rotational temperature for all sources where CO 10--9 data are available; similar figures for individual sources are available in the Appendix (Fig. \ref{fig:trot_ind}). The mean ratio averaged over all sources and velocities is 0.7, corresponding to a rotational temperature of $\sim$ 350 K. Most sources show a slightly lower ratio, by up to a factor of two, closer to the line center. This trend suggests that the gas at the higher velocities is warmer than at lower velocities, similar to what is seen toward outflow spots well offset from the source position \citep[e.g., L1157-B1,][]{lefloch12}. If a linear scaling is applied to account for different beam sizes, the mean ratio decreases to 0.4 (240 K), and if a point-source scaling is applied the ratio decreases to 0.2 (180 K). 

\textit{Herschel}-SPIRE and PACS measure the integrated intensity of CO emission from $J$=4--3 to 49--48, and any rotational temperature inferred from these observations are therefore intensity-weighted averages of the rotational temperatures calculated here. In order to do a direct comparison, the velocity-dependent intensities were used as weights for determining the average rotational temperatures. These intensity-weighted temperatures tend to be lower by 20 K (emission fills the beam) to 10 K (point source scaling), and the effect is therefore negligible.

\subsection{H$_2$O and CO}\label{sect:h2o}

\textit{Herschel}-PACS footprint maps of CO 16--15 and various H$_2$O transitions show that both species follow one another spatially, as opposed to for example CO 3--2 and H$_2$O which are spatially and kinematically distinct \citep[e.g.,][]{santangelo12}. This spatial correlation suggests that the best reference frame for measuring H$_2$O abundances is a high-$J$ CO line such as the $J$ = 16--15 transition. 

In the following analysis, the H$_2$O abundance is measured from both the 1$_{10}$--1$_{01}$ transition at 557 GHz and the 2$_{12}$--1$_{01}$ transition at 1670 GHz (179.5 $\mu$m). The former transition is the only one observed toward all sources, but in a 38$''$ beam \citep{kristensen12, green13}. The latter, on the other hand, is only observed toward six sources (Table \ref{tab:sample}) but in a 13$''$ beam \citep{mottram14}. Thus, this line may be used to calibrate the abundance inferred from the 557 GHz transition. The ratio of these two lines is nearly constant as a function of velocity for all six sources, suggesting that both lines are equally good for calculating the water abundance, and also that most of the water emission originates from within the smaller 13$''$ beam \citep{mottram14}, but possibly with weaker emission extending beyond the beam. This weak emission is indeed seen in \textit{Herschel}-PACS footprint maps \citep[Table \ref{tab:sample},][]{karska13, green13}. 

\begin{figure}[!t]
\begin{center}
\includegraphics[width=\columnwidth]{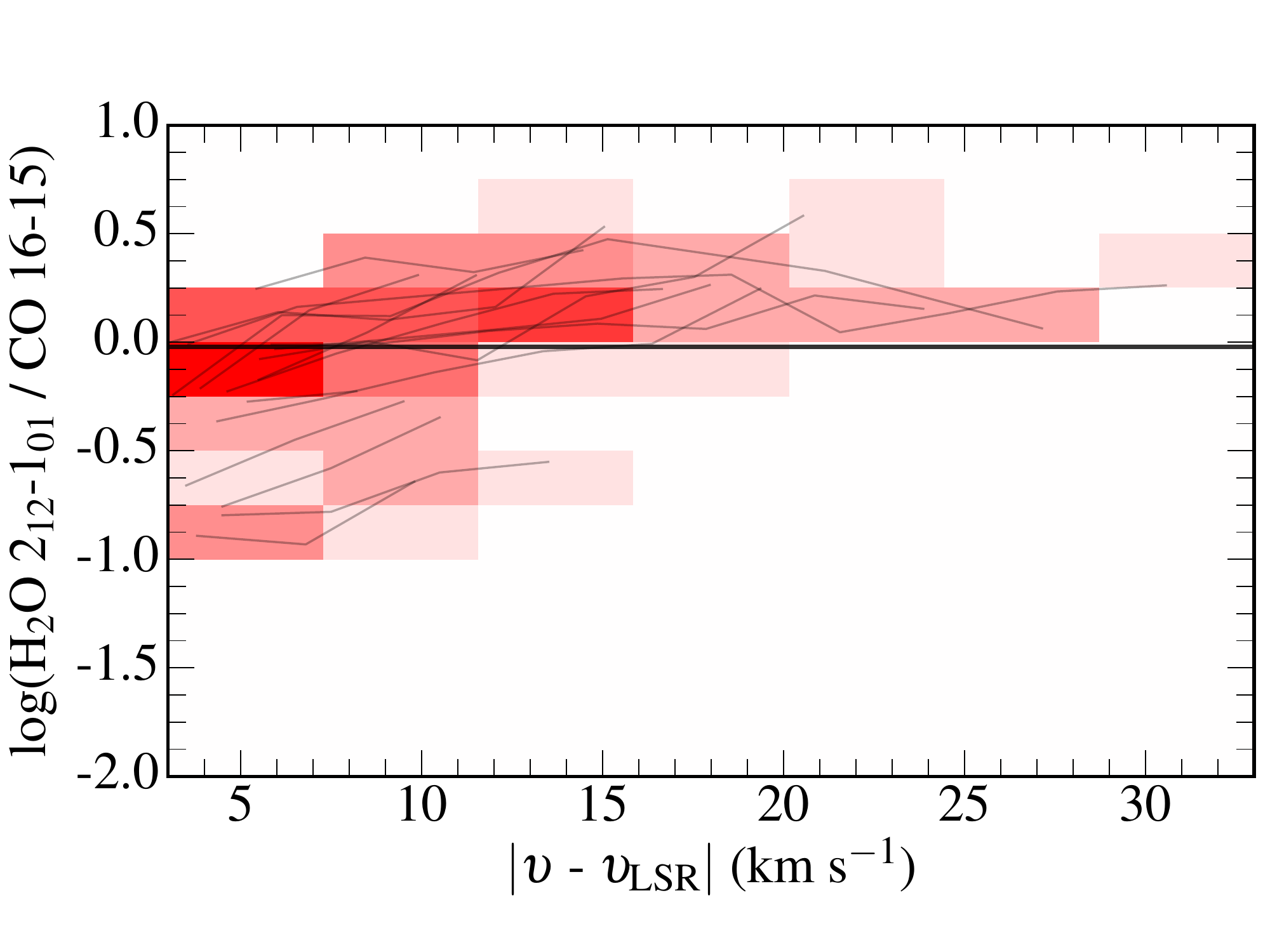}
\includegraphics[width=\columnwidth]{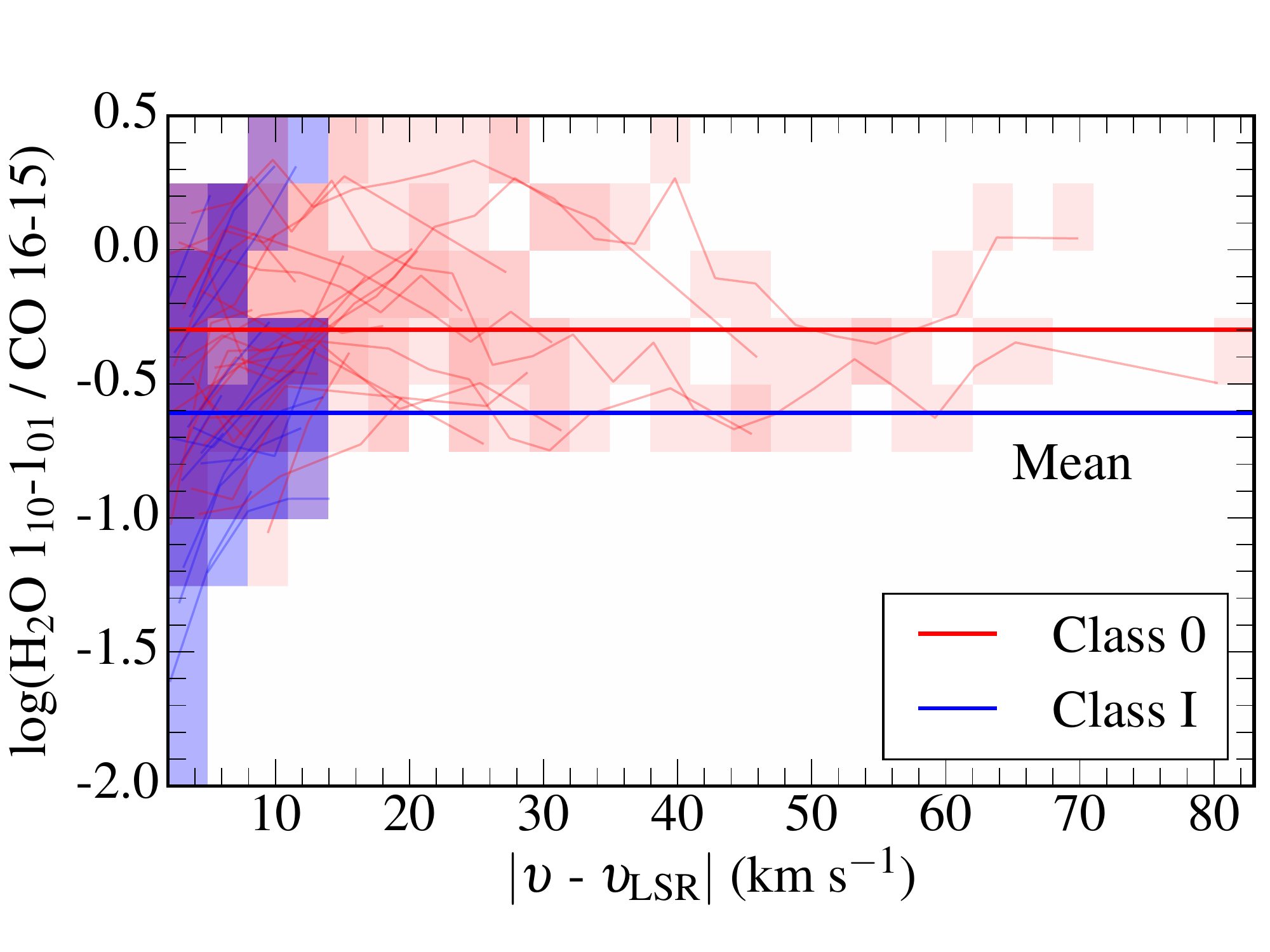}
\caption{\textit{Top:} H$_2$O 2$_{12}$--1$_{01}$ / CO 16--15 line ratio as a function of velocity toward the six Class 0 sources where the H$_2$O line is observed with HIFI. The brightness of the color indicates the number of observations passing through a given point---brighter represents a higher number. The channel size is 3 km\,s$^{-1}$. \textit{Bottom:} Same as \textit{top} but for the H$_2$O 1$_{10}$--1$_{01}$ / CO 16--15 line ratio where Class 0 and I sources are separated (red and blue, respectively). The total mean ratio is displayed. 
\label{fig:h2o_ratio}}
\end{center}
\end{figure}

The H$_2$O 2$_{12}$--1$_{01}$ / CO 16--15 line ratio is shown in Fig. \ref{fig:h2o_ratio}. The central $\pm$2 km\,s$^{-1}$ are masked out as the H$_2$O line suffers from deep self absorption associated with cold envelope material \citep[Fig. \ref{fig:smm1_co},][]{mottram14}. Furthermore, both spectra are rebinned to 3 km\,s$^{-1}$ channels to increase $S/N$ in the line wings. The line ratio shows a tendency to rise from 0.5 (low velocity) to 1.7 (high velocity) but the ratio is consistent with being constant at 1.2$\pm$0.8, where the uncertainty represents the spread in values. In the following the ratio will be treated as constant. 

Similarly, the H$_2$O 1$_{10}$--1$_{01}$ / CO 16--15 flux ratio is constant as a function of velocity with an average value of 0.42$\pm$0.36 (Fig. \ref{fig:h2o_ratio}). If a linear scaling is applied to account for the differences in beam size, the ratio increases to 1.5$\pm$1.3, and it further increases to 5.3$\pm$4.6 for a point-source scaling. There is no significant difference between Class 0 and I sources, although Class I sources tend to show a slightly lower line ratio. See Fig. \ref{fig:h2o_ind} in the Appendix for individual sources. 

To translate the line ratios into abundance ratios, or, more appropriately, column density ratios, a set of non-LTE 1D radiative-transfer models are run. The code {\sc Radex} is used \citep{vandertak07} with collisional rate coefficients from \citet{yang10} and \citet{neufeld10} for CO, and \citet{daniel11} for H$_2$O, as tabulated in LAMDA \citep{schoier05}. The \textsc{Radex} code was modified according to \citet{mottram14} to account for the high opacity of the H$_2$O transitions. The excitation conditions (density and H$_2$O column density) are taken from \citet{mottram14} and a temperature of 300 K is used together with $\Delta V$=20 km\,s$^{-1}$. \citet{mottram14} found no deviation from an H$_2$O ortho/para ratio of 3, the high-temperature statistical equilibrium value, and that is used here. Based on {\sc Radex} modeling, \citeauthor{mottram14} found that the best-fit solutions fell in two camps: subthermal excitation of the H$_2$O transitions with $n$(H$_2$)=10$^6$ cm$^{-3}$, $N$(H$_2$O)=4$\times$10$^{16}$ cm$^{-2}$, and thermal excitation with $n$(H$_2$)=5$\times$10$^7$ cm$^{-3}$, $N$(H$_2$O)=10$^{18}$ cm$^{-2}$. Finally, \citeauthor{mottram14} used a linear beam scaling, and that is applied here as well. The only free parameter is the CO column density which is varied until the two H$_2$O / CO line ratios are reproduced. 

\begin{figure}[!t]
\begin{center}
\includegraphics[width=\columnwidth]{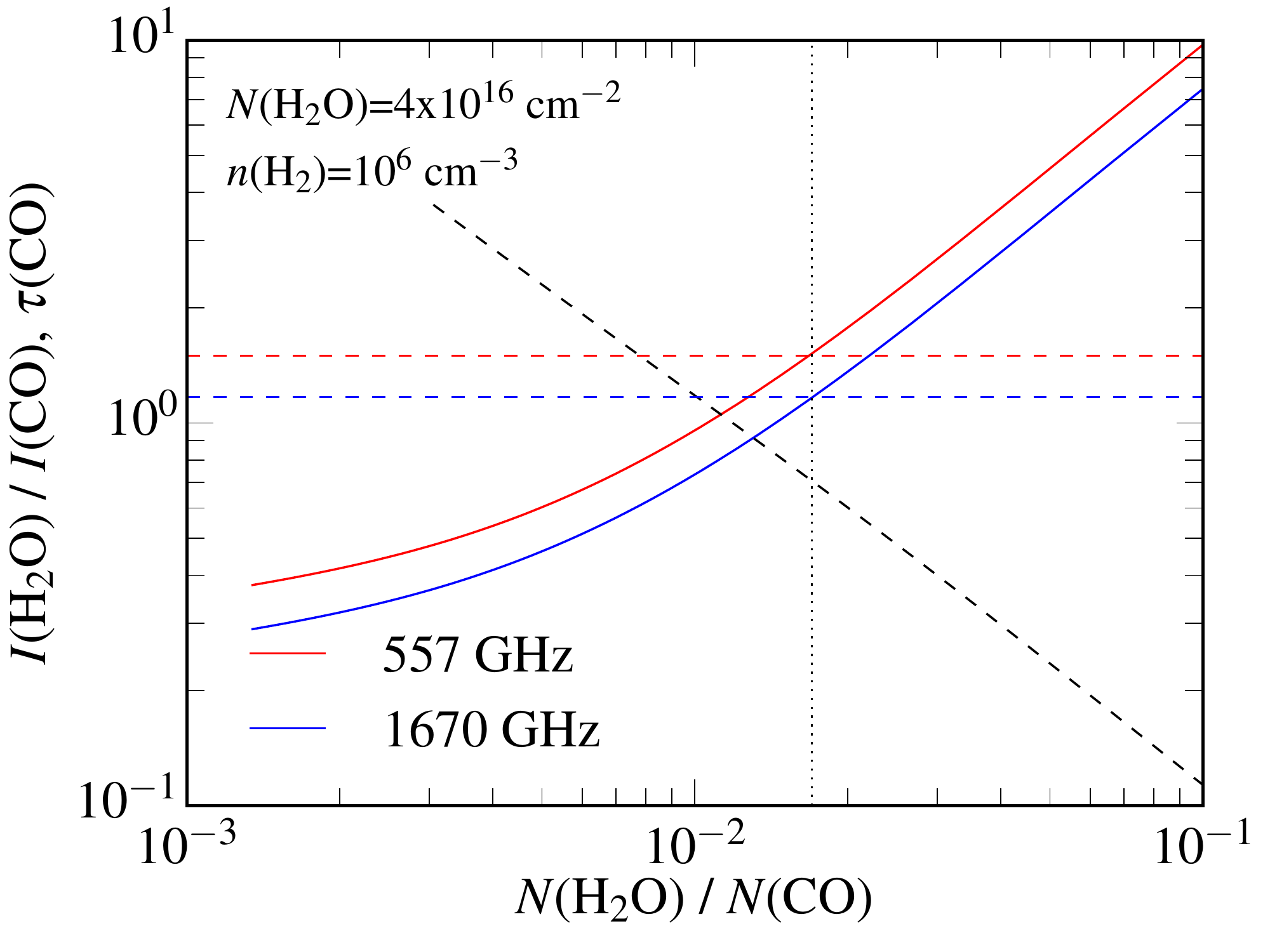}
\includegraphics[width=\columnwidth]{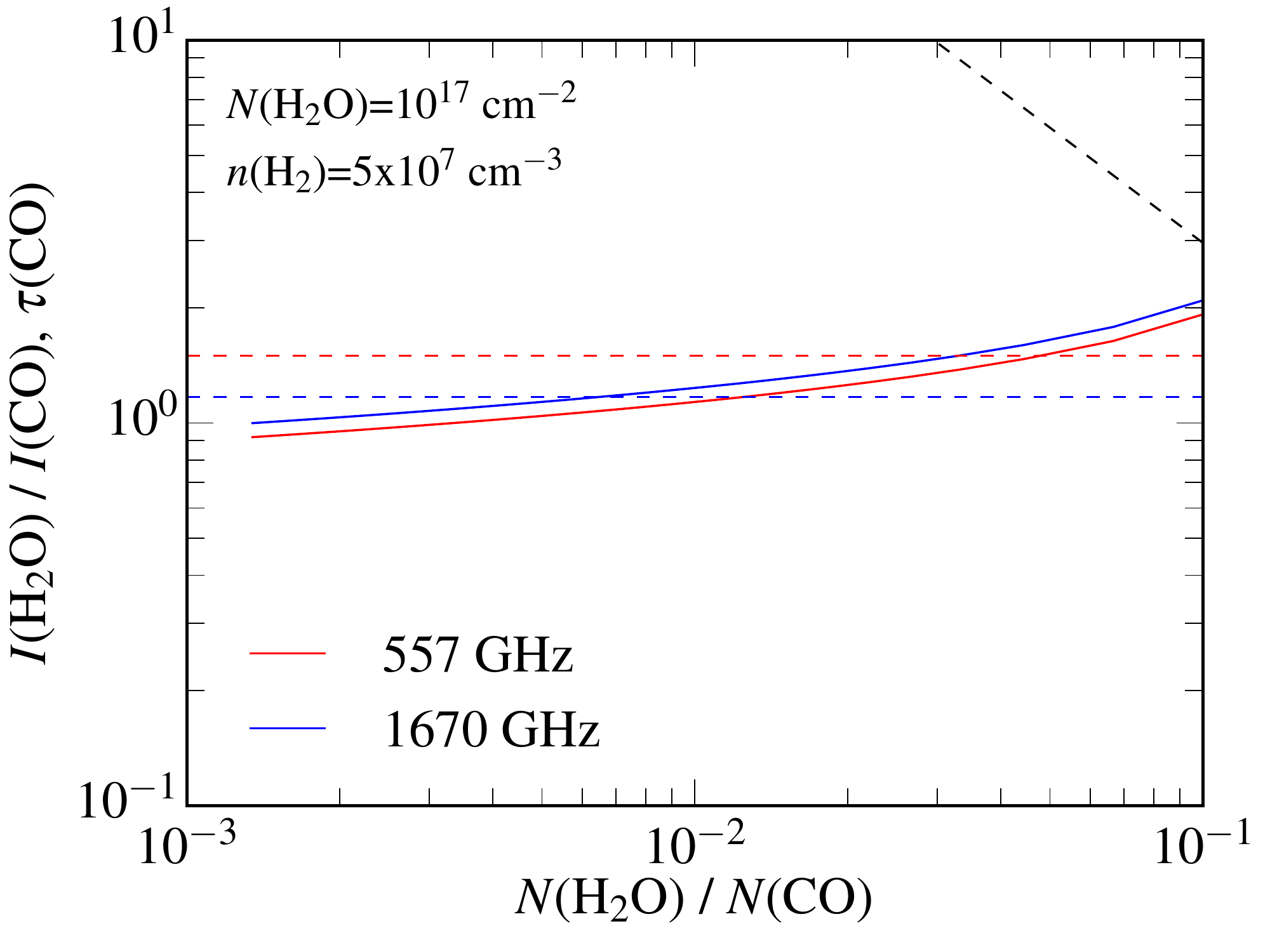}
\caption{Modeled intensity ratio of the 557 and 1670 GHz H$_2$O lines over CO 16--15 as a function of the H$_2$O / CO column density ratio, where the H$_2$O column density is both ortho- and para-H$_2$O. The observed ratios are shown as dashed horizontal lines. The black dashed line is the CO 16--15 opacity, and is shown on the same scale as the intensity ratio. \textit{Top:} subthermal excitation conditions for H$_2$O. \textit{Bottom:} Thermal excitation conditions for H$_2$O. 
\label{fig:h2o_abun}}
\end{center}
\end{figure}

Figure \ref{fig:h2o_abun} shows the modeled intensity ratios of the \mbox{H$_2$O 2$_{12}$--1$_{01}$} and 1$_{10}$--1$_{01}$ lines to CO 16--15 versus the total H$_2$O / CO column density ratio for both subthermal and thermal H$_2$O excitation, together with the opacity of the CO 16--15 transition. For subthermal H$_2$O excitation conditions, a column density ratio of H$_2$O/CO of 0.02 best reproduces both line ratios, corresponding to a total H$_2$O abundance of 2$\times$10$^{-6}$ for a CO abundance of 10$^{-4}$ \citep{dionatos13}. For the thermal excitation conditions, a large range of H$_2$O/CO column density ratios reproduce emission, ranging from $N$(H$_2$O)/$N$(CO)$\sim$10$^{-3}$ to 10$^{-1}$. However, all these solutions have one thing in common: CO 16--15 must be optically thick with $\tau$ $>$ 3 over the entire profile. Toward the source with the highest CO 16--15 flux in the sample, Ser SMM1, the observed $^{12}$CO 16--15 / $^{13}$CO 16--15 line ratio is 55 \citep{goicoechea12}, consistent with the $^{12}$CO line being optically thin. It is therefore unlikely that any other source shows optically thick emission, and the thermal H$_2$O excitation solution can be excluded.

\section{Discussion}\label{sect:disc}

\subsection{CO excitation}

CO line profiles change gradually with excitation, ranging from the relatively narrow, centrally peaked CO 2--1 and 3--2 profiles typically observed toward outflow regions, to the broader high-$J$ CO lines; $\varv_{\rm max}$ remains constant (Fig. \ref{fig:smm1_co}). This gradual change suggests that these CO 16--15 line profiles reveal a high-temperature component not previously observed from the ground. The change in profile shape cannot be attributed to opacity effects, apart from a few km\,s$^{-1}$ around the source velocity. The opacity will be lower for higher-excitation lines, and the emission at the lower velocities should therefore increase more compared to higher-velocity material if opacity played a major role. Yet the opposite trend is observed: emission at higher velocities increases with respect to emission at lower velocities. Opacity, therefore, plays a minor role in setting the line profile for CO 16--15. 

The gradual change in CO line profile with excitation suggests that different CO transitions, up to at least $J$ = 16--15, trace material at different temperatures, regardless of whether CO is thermally or subthermally excited. If all CO emission were to trace a single-temperature, single-density slab, the profiles should be identical as is the case for water line profiles \citep{mottram14}. The rotational temperature of the emitting material changes as a function of velocity (Sect. \ref{sect:trot} and Fig. \ref{fig:trot}), higher temperatures at higher velocities, consistent with what is seen at outflow spots \citep{lefloch12} and the protostellar position in lower-$J$ transitions \citep{yildiz13}. A single isothermal slab is therefore not sufficient for reproducing the CO excitation. 

When taking the most conservative estimate of the rotational temperature, 180 K obtained from the point-source scaling, the temperature is significantly higher than found for lower-$J$ CO transitions \citep[median $T_{\rm rot}$$\sim$70 K for transitions up to 10--9;][]{vankempen09, goicoechea12, yildiz13, yang17}. The rotational temperature does approach the typical rotational temperature of $\sim$ 300 K often found for this part of the CO ladder toward embedded protostars \citep{manoj13, karska13, green13}. 

The actual kinetic temperature of the CO 16--15 emitting gas is higher than the rotational temperature measured here. The part of the CO rotation diagram going up to $J$ $\sim$ 14 shows positive curvature \citep{goicoechea12, yildiz13}, and from $J$ $\sim$ 14 to 25 the ladder shows little curvature \citep{manoj13, green13}. This curvature implies that a rotational temperature measured from, say, the $J$ = 15--14 and 16--15 transitions will be higher than when measured from the $J$ = 10--9 and 16--15 transitions and likely closer to that observed with PACS in the warm 300 K component. Thus, only a fraction of CO 10--9 emission traces the broad cavity shock component seen so prominently in CO 16--15 and H$_2$O profiles. 

The relative contributions to the HIFI line profiles may be compared to the warm ($T_{\rm rot}$ $\sim$ 300 K) and hot ($T_{\rm rot}$ $\sim$ 800 K) components observed with PACS \citep[][Karska et al. in prep.]{manoj13, karska13, green13}. The break between the two PACS components typically appears around $J$=25--24 ($E_{\rm up}$ = 1800 K) for low-mass protostars. Given the rotational temperatures and break point, if the two rotational temperatures have different physical origins then the warm and hot PACS components contribute on average $\sim$ 80\% and 20\% to the total CO 16--15 flux, respectively. These PACS fractions do not depend on evolutionary stage or any other source property (Karska et al. in prep.), but the fractions are constant when both components are detected. Any contribution from the very hot ($T_{\rm rot}$ $>$ 1000 K) component will not be detected in the CO 16--15 line profiles, and it is ignored in the following. 

The PACS fractions depend on an extrapolation from $J_{\rm up}$~$\ge$~25 ($E_{\rm up}$/$k_{\rm B}$ $\sim$ 1800 K) down to $J$=16--15 ($E_{\rm up}$/$k_{\rm B}$ $\sim$ 750 K). Such an extrapolation carries uncertainties, especially if not all high-$J$ CO lines are observed or detected. The extrapolation uncertainty was inferred for two representative sources, Ser-SMM1 and B335. The former is one of the sources with the highest $S/N$ on the high-$J$ CO transitions observed with PACS, whereas the latter was only detected in a limited number of transitions and at low $S/N$. For Ser-SMM1 the extrapolation uncertainty is 4\%, while it is 11\% for B335. Some additional uncertainty may arise from how well-centered the sources are on the central PACS spaxel, and we therefore adopt a typical extrapolation uncertainty for all sources of 10\%. 

If the CO ladder consists of two temperature components, and thus two physical components, then 80\% of the flux originates in the warm 300 K component and 20\% in the hot 750 K component. The CO 16--15 HIFI line profiles typically consist of multiple components; the dominant flux component is the broad cavity shock component, which typically contributes 75\%$\pm$20\% to the total flux. These relative fractions clearly overlap within the uncertainty. This overlap, and the measured rotational temperature from CO 10--9 and 16--15 (200--300~K), strongly suggests that the PACS components may be associated with corresponding HIFI components, i.e., the warm, 300 K PACS component may be associated with the broad outflow cavity shock HIFI component, and the warm 750 K PACS component to anything else contained in the HIFI line profiles. These latter components only appear in CO 16--15 profiles and not at lower $J$, clearly suggesting these components arise in hot gas. Furthermore, the multiple components in the HIFI line profiles implies that the PACS CO ladder consists of multiple physical components.

\subsection{Physical origin of CO emission}

The high-$J$ CO emission can arise in three regions within the protostellar system: the disk; the UV-heated cavity walls; the outflow, or a combination thereof. In this context, the term outflow encompasses the wind, the envelope entrainment layers, and the jet. None of the profiles show double-peaked features as would be expected if the emission arises from the disk surface \citep{bruderer12, fedele13, harsono13}, and so disks are quickly ruled out as a dominant source of the observed CO emission. UV-heated outflow cavity walls are ruled out because the 300 K PACS component is ubiquitous: the protostars observed here, and in other samples, span a range of three orders of magnitude in bolometric luminosity ($\sim$0.1--100 $L_\odot$). The UV luminosity is expected to span a similar range. For this reason, and as argued by \citet{manoj13}, it seems unlikely that UV heating plays a dominant role in generating this emission because the UV luminosity of each source would need to be specifically tuned to reproduce a rotational temperature of 300 K everywhere. Thus, the dominant excitation mechanism must be outflow-related.

\begin{figure*}[!t]
\begin{center}
\includegraphics[width=14cm]{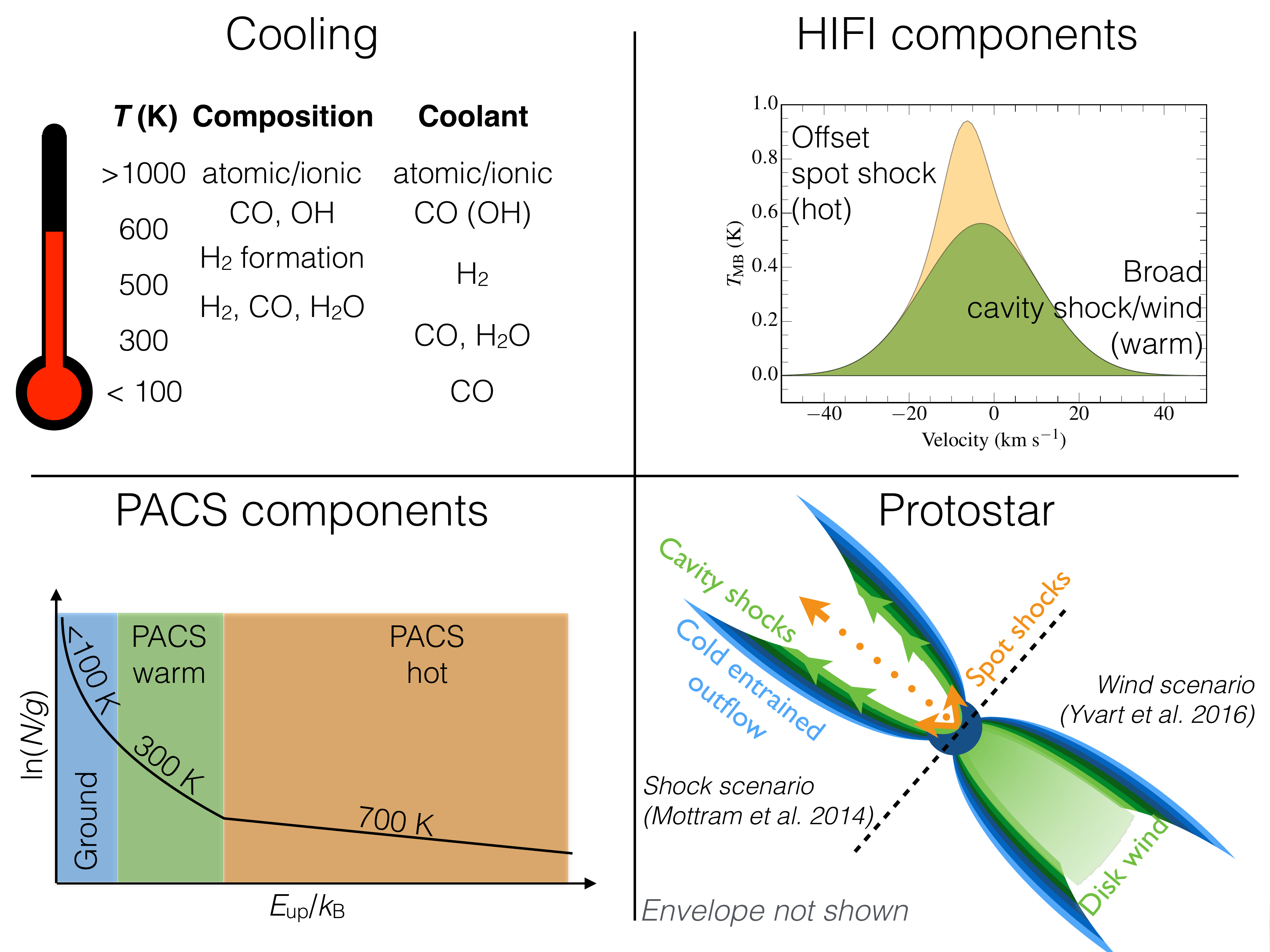}
\caption{Schematic showing the different parts to the proposed interpretation. The top left panel shows the cooling of atomic to molecular gas, along with the dominant constituents and coolants at each stage. Top right shows a cartoon HIFI CO 16--15 profile centered at zero velocity, with the broad outflow cavity shock component shown in green, and the offset spot shock component in orange. Bottom left is a CO rotational diagram with the different temperature components highlighted. Finally, bottom right shows where these different components may be located in the protostellar system. The top part shows the scenario where the broad component is caused by outflow cavity shocks, and the lower part shows the scenario where the broad component arises in the disk wind. 
\label{fig:cartoon}}
\end{center}
\end{figure*}

\subsubsection{Origin of warm PACS emission}

The bulk of the CO 16--15 emission is related to outflowing gas, both for Class 0 and I sources. Based on the association with water, it seems likely that the CO 16--15 emission primarily traces gas which is currently interacting with, or passing through, shocks, as opposed to the slower, colder entrained outflow traced by the lower-$J$ CO emission \citep[see also][]{kristensen12, mottram14}. The radius of the water-emitting region, and therefore also the radius of the CO 16--15 emitting region, is typically of the order of 10$^2$ AU \citep{mottram14}. If the emitting region is not circular but rather cylindrical, a cylinder with a diameter of 15--30 AU could account for the linear geometrical beam scaling found to be appropriate in Sect. 3.4. Such a width is consistent with both the H$_2$O and CO cooling lengths of continuous C-type shocks with pre-shock density $\geq$~5$\times$10$^5$~cm$^{-3}$, almost irrespective of shock velocity \citep{kristensen07, visser12}. In this scenario, the cavity shocks are thus located along the outflow cavity walls in a thin ($\lesssim$~10~AU) layer \citep{mottram14}. 

An alternative to this scenario is that this component traces a molecular disk wind \citep{panoglou12, yvart16}. If the wind is accelerated at a steady pace, it will keep its molecular content, particularly during the deeply embedded stages where the dusty wind shields the gas from dissociating UV photons from the accreting protostar. \citet{yvart16} used this model to reproduce the H$_2$O 1$_{10}$-1$_{01}$ spectra presented in \citet{kristensen12} and the excited line profiles in \citet{kristensen10}, and were able to reproduce the width and intensities of the profiles with only two free parameters: the mass accretion rate and inclination angle. Clearly this model presents a very attractive alternative to the origin of water emission, and potentially also to high-$J$ CO emission. An important test will be to see how well the model reproduces not only the H$_2$O spectra but also the CO 16--15 spectra presented here, as well as emission from the entire CO ladder. 

Irrespective of the underlying physical origin, an important question remains: why is the 300 K component ubiquitous in the PACS data with very little scatter, independent of physical conditions? What is it in these broad cavity shocks seen in the CO 16--15 line profiles that generates this temperature component? The fact that the rotational temperature is ubiquitous must imply that some fundamental aspect related to the gas cooling controls the excitation, since one would otherwise expect the excitation to depend on local parameters. Particularly, the observed dominant gas coolants \citep[H$_2$, CO, H$_2$O, OH and O;][]{karska13} are responsible for setting the overall temperature structure or distribution of the gas. A species may dominate the cooling over a specific range of temperatures and densities if that species is either particularly abundant, or particularly efficient at cooling the gas through excitation/deexcitation effects. If the species dominates the cooling through excitation/deexcitation effects, it will do so because the level populations react more efficiently to the change in temperature than those of other species. This information is relayed to other species through collisions. Testing this hypothesis will be done through more detailed calculations of the cooling functions of the observed dominant coolants, that includes calculating their level populations explicitly in the cooling gas (Kristensen \& Harsono in prep.).

The second part of the above question is: why always 300 K? One possibility is that the other dominant molecular coolant of warm/hot gas, H$_2$, stops being an efficient coolant around 300 K because of the widely-spaced energy levels (the $J$ = 2--0 transition has $E_{\rm up}$/$k_{\rm B}$ = 510 K), i.e., because other species become more efficient at cooling the gas through excitation effects. Indeed, such a scenario would explain why CO cooling always activates at this kinetic temperature, as has been shown to be the case in shocks \citep{neufeld89, flower10, flower15}, see Fig. \ref{fig:cartoon}. Once H$_2$ ceases to be effective, CO, as the new dominant coolant, brings the temperature gradually down to ambient temperatures. 

\begin{figure}[!t]
\begin{center}
\includegraphics[width=\columnwidth]{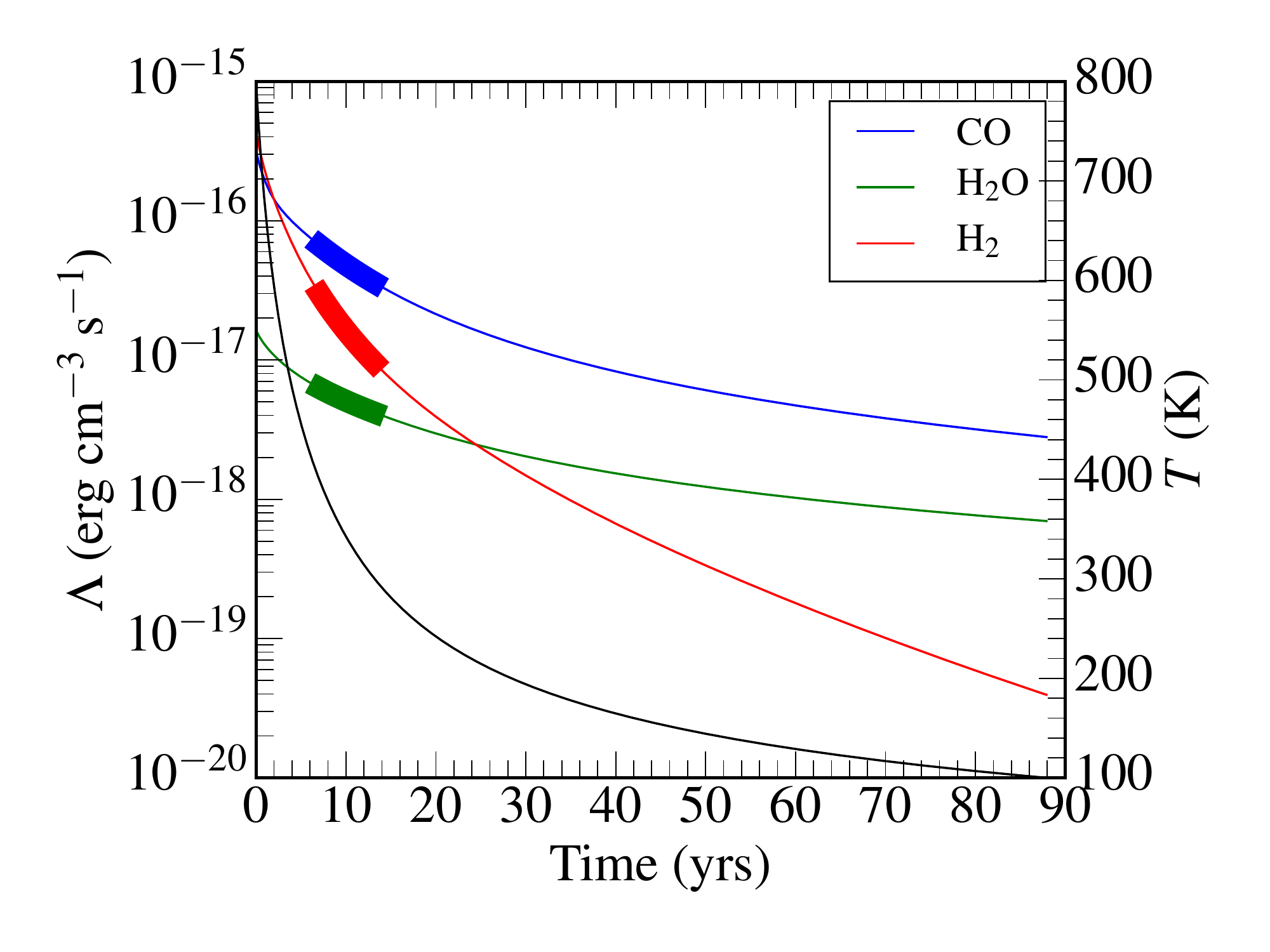}
\caption{Cooling curves for H$_2$, CO, and H$_2$O using the analytical cooling functions of \citet{neufeld93}. The colored lines are the cooling rates, and their values are shown on the left axis. The black line is for temperature (right axis). The highlighted parts on the cooling curves are for the temperature range 300--400 K. The H$_2$ density and column densities are as shown in Fig. \ref{fig:h2o_abun}. \label{fig:neufeld}}
\end{center}
\end{figure}

Such a scenario can be tested by utilizing existing analytical cooling functions \citep{neufeld93}, the results of which are shown in Fig. \ref{fig:neufeld}. Here, cooling rates are calculated for H$_2$, CO, and H$_2$O for a gas with density $n$ = 10$^6$ cm$^{-3}$, $N$(H$_2$O)/$N$(CO) = 0.02, and starting temperature of 1000 K. In this simple calculation, CO takes over as the dominant coolant at $\sim$ 500 K, i.e., somewhat higher than the observed cross-over temperature of $\sim$ 300--400 K. We note that these analytical cooling functions are calculated for collisions with H$_2$, i.e., there is no coupling to chemistry, particularly H$_2$ formation. The more detailed calculations currently underway will provide more insight into how the level populations actually respond to the changes in the cooling gas, both physical and chemical, and over which timescales.

These analytical cooling functions generally show three regimes in which species dominate cooling: at low density ($<$ 10$^5$ cm$^{-3}$), H$_2$ dominates cooling to the lowest temperatures ($<$ 100 K) because CO is not efficiently excited. At high densities ($>$ 10$^8$ cm$^{-3}$), CO is the dominant coolant throughout because the level populations are very efficiently populated all the way to very high $J$. In the intermediate regime, which is where H$_2$O and CO 16--15 are excited \citep{mottram14}, there is a cross-over point where H$_2$ starts as the dominant coolant, and CO then takes over. 

\subsubsection{Origin of hot PACS emission}\label{sect:hot}

The division between warm and hot CO emission implies that the hot 700 K PACS component is traced by anything else in the line profiles. For the Class 0 sources, this typically means spot shocks where the wind from the protostar impinges on the inner cavity wall. The pre-shock gas is likely dissociated to some degree by the UV radiation from the accreting protostar, and the shock itself is relatively slow \citep[5--10 km\,s$^{-1}$,][]{kristensen13, benz16}. This conclusion was reached based on detection of light ionized hydrides in this particular component, the detection of the offset spot shock component in H$_2$O and CO 16--15 but not lower-$J$ transitions, and the component kinematics (offset velocity and $FWHM$). These three observational facts (kinematics, chemistry, and excitation of the component) point to an origin in a $J$-type shock. 

The spot shocks and narrow components typically peak at blue-shifted velocities from the source velocity. If the protostellar wind impinges on the inner envelope walls at very small radii ($\lesssim$ 100 AU), the optically thick protostellar disk may shield the part of the wind moving away from us, i.e., the red-shifted part of the component \citep{alma15, kristensen13, kristensen16}. If, on the other hand, the shocks interact with the inner envelope on scales larger than the disk, or if the disk is not optically thick at these wavelengths, both sides will be seen and the profile will appear more symmetric. Such partial continuum shielding is similar to what has been inferred from optical atomic-line observations of disk winds, where the line profiles appear skewed \citep[e.g.,][]{edwards87}. 

The lower velocity shifts observed toward Class I sources compared to Class 0's are consistent with the protostellar wind velocity decreasing radially away from the jet axis \citep{panoglou12}. As the outflow cavity opening angles increase with evolutionary stage, the part of the wind interacting with the envelope will decrease in velocity. However, only a very small velocity is required to produce the high temperature observed in the high-$J$ part of the CO ladder. For J-type shocks, the peak temperature is 
\begin{equation}
T_{\rm peak} = \frac{3}{16} \frac{\mu \varv^2_{\rm shock}}{k_{\rm B}}\, ,
\end{equation}
where $\mu$ is the mean molecular mass, $\varv_{\rm shock}$ the shock velocity, and $k_{\rm B}$ is Boltzmann's constant. To reach a peak temperature of 1000 K, a shock velocity of only $\sim$ 5 km\,s$^{-1}$ is required if the pre-shock gas is atomic. The above relation only applies to strong shocks where the shock speed greatly exceeds both the sound speed and the Alfv{\'e}n speed, and thus may not be fully applicable to the conditions near a protostar. However, the calculation still serves to show that even shocks moving at moderate velocities are enough to heat the gas. We furthermore note that the observations likely do not probe a single shock with a single velocity, but that shocks with a range of shock velocities is contained in the beam. 

The scenario is as follows: close to the protostar and on the side of the outflow cavity wall facing the protostar, UV photons dissociate gas and it is either neutral atomic, or ionized. As the slow wind impinges on this gas, it heats it to temperatures $>$~1000~K, and compresses it (Fig. \ref{fig:cartoon}). The compression leads to partial shielding from the UV irradiation, and the heating initiates a shock chemistry, similar to that explored by \citet{neufeld89} and summarized here. In this chemistry, CO and OH form first and these two species are the dominant coolants until H$_2$ formation sets in. H$_2$ formation becomes efficient at $\sim$ 500 K, and as the H$_2$ molecules form, the gas temperature remains constant at $\sim$ 500 K, heated by the liberated binding energy. As the gas continues to compress, shield, and cool, H$_2$O forms efficiently from the now abundant H$_2$ and OH (plus any remaining atomic O not locked up in OH). When the temperature decreases to $\sim$ 300 K, H$_2$ stops being the dominant coolant, and CO and H$_2$O take over, as these are the most abundant coolants. Emission in the narrow, offset shock component is thus dominated by emission from the hot pre-H$_2$ formation zone, whereas H$_2$O emission is dominated by the post-H$_2$ formation zone: the H$_2$O-emitting part is located further into the envelope where the shock velocity has already been dampened. This coherent picture provides a natural explanation for the observed characteristics of the narrow component, and the interpretation is similar to what what proposed for the spot shock identified by \citet{kristensen13}. However, further model calculations are required to verify this picture. 

The narrower components do not appear to be detected in lower-$J$ CO emission all the way up to CO 10--9, consistent with them tracing hotter material. UV-heated cavity walls have been uniquely identified in low-$J$ $^{13}$CO emission, specifically \mbox{$^{13}$CO 6--5} \citep{yildiz12}. The lack of any contribution to the high-$J$ CO ladder suggests that the UV-heated material is colder, and that the temperature distribution with mass must be steeper than in shocks, i.e., there is less material at higher temperatures relative to lower temperatures than in a shock. This is consistent with the temperatures inferred by \citet{yildiz15}, where the temperature of the UV-heated gas is of the order of $\sim$ 50 K, i.e., it would not contribute strongly to the CO 16--15 emission. 

The above proposed scenario depends on one factor: H$_2$ dissociating UV photons must be present on small scales. If not, the shock speeds must be high enough that the shocks themselves are either strong enough to dissociate H$_2$, $\varv$ $\gtrsim$ 30 km\,s$^{-1}$, or strong enough to produce the H$_2$-dissociating UV photons, $\varv$~$\gtrsim$~50~km\,s$^{-1}$ \citep{neufeld89}. Both these shock velocities are too high to account for the narrow emission profiles and the velocity offsets. The presence of the dissociating photons is consistent with the scenario presented by \citet{visser12} and \citet{lee14} in that H$_2$O molecules are dissociated in the UV-heated cavity walls. To what extent UV photons aid in further heating of the cavity shocks is an open question; it is evident that the UV photons still play a crucial role in the feedback from accreting protostars \citep{hull16} and in setting the chemistry \citep{kristensen13, benz16}, although not necessarily in the excitation of the $^{12}$CO ladder. 

\subsection {H$_2$O abundance}\label{sect:disc_h2o}

The inferred H$_2$O abundance is two orders of magnitude lower than expected if all non-refractory oxygen not locked up in CO is incorporated in water, and if all water is released from the icy grain mantles \citep[3$\times$10$^{-4}$;][]{vandishoeck14}. This result applies not only to the source position, but the water abundance is generally observed to be low also toward off-source outflow positions \citep[e.g.,][]{bjerkeli12, tafalla13, santangelo13}. Three possible explanations exist for this lack of water: \textit{(i)} water remains on the grains or quickly re-adsorbs after the passage of a shock wave; \textit{(ii)} the warm gas-phase synthesis is slowed down or reversed by large amounts of atomic hydrogen; or \textit{(iii)} UV photons dissociate water. These three possibilities will be discussed further below. 

Icy grain mantles are sputtered through impacts with H, H$_2$, and He in shock waves where the charged and neutral fluids are streaming past each other. Shock models demonstrate that complete removal of the icy mantles happens at $\varv$ $\gtrsim$ 10 km\,s$^{-1}$ \citep[e.g.,][]{flower10}. Another independent line of study demonstrates that in outflows, the observed column density ratio of the two grain-species water and methanol is similar to that observed in ices at the lowest outflow velocities \citep[$\lesssim$~10~km\,s$^{-1}$;][]{suutarinen14}, but they diverge at higher velocities where methanol is collisionally dissociated. Thus the icy grain mantles are likely completely removed in outflows. The freeze-out time estimated in \citet{bergin98} is typically $>$~10$^5$~yrs, i.e., significantly longer than the dynamical lifetime of the outflows; if the density is higher than 10$^5$ cm$^{-3}$, the freeze-out time may be correspondingly shorter. However, it remains unlikely that of order 99\% of all water hides on grains, as would be required to explain the observed line ratios. 

If H$_2$ is dissociated, either collisionally or through photodissociation, the reverse reaction (H$_2$O $+$ H $\rightarrow$ OH $+$ H$_2$) may dominate and collisionally dissociate water at high temperatures. The bond energy of the O-H bond in water is similar to the H-H bond in H$_2$ (5.11 versus 4.52 eV) and so it is difficult to imagine a situation in which H$_2$ is preferentially collisionally dissociated without also dissociating water, and this combination would rapidly lead to complete dissociation of all water. Photodissociation of H$_2$ occurs at discrete wavelengths in the range 912--1050~\AA, whereas H$_2$O is photodissociated over the entire wavelength range of 912--2000~\AA\ \citep{fillion01}, implying that water is more readily photodissociated. Moreover, H$_2$ starts to self-shield even at very low $A_V$, further lowering photodissociation of this molecule. Thus, water is more readily photodissociated than H$_2$, and it seems unlikely that any process will enhance the H abundance without destroying water at the same time. An important test of this scenario will be to observe these systems with high sensitivity in the H I 21cm line with, e.g., the JVLA, and measure the abundance of atomic H with respect to CO and H$_2$O, as has previously been done for a few systems \citep[e.g.,][]{lizano88, rodriguez90, lizano95}. 

\begin{figure}[!t]
\begin{center}
\includegraphics[width=\columnwidth]{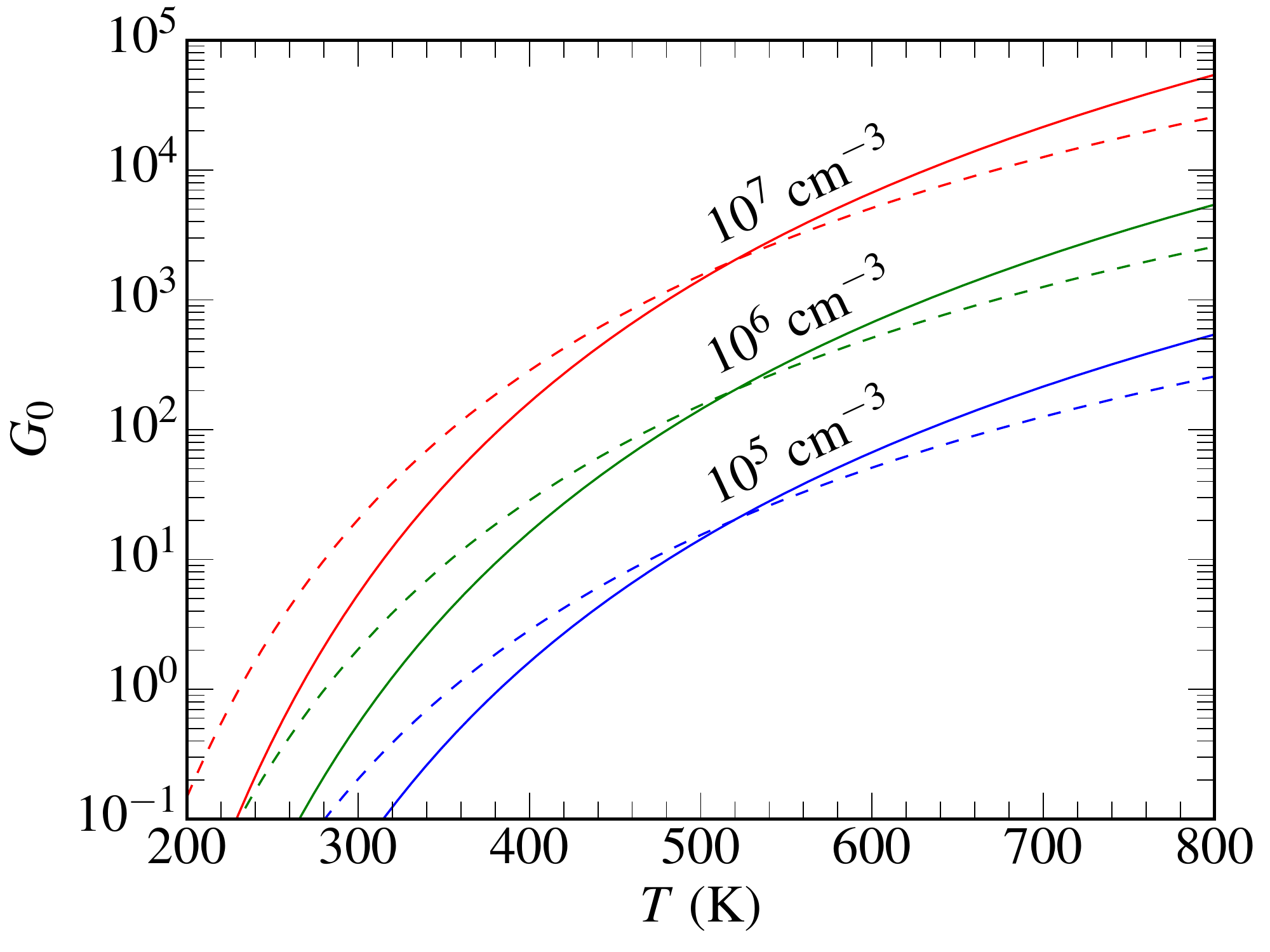}
\caption{Required value of $G_0$ for three different H$_2$ densities, 10$^5$, 10$^6$, 10$^7$ cm$^{-3}$ as a function of gas temperature to reach an equilibrium H$_2$O abundance of 2$\times$10$^{-6}$, assuming photodissociation is the only path of destruction. The results from the network including OH from \citet{bethell09} are shown with dashed lines for the same densities. 
\label{fig:h2o_diss_form}}
\end{center}
\end{figure}

Finally, water may be photodissociated directly, even in environments where the H$_2$ abundance is high. To estimate the required intensity of the UV field, in units of the interstellar UV field, $G_0$, two assumptions are made: first, all water that is photodissociated immediately dissociates all the way to atomic oxygen, either directly or in two steps with OH photodissociation following H$_2$O photodissociation; second, the only water formation route is through gas-phase synthesis with H$_2$ at high temperature. For a water abundance of 10$^{-6}$, the atomic oxygen abundance is 10$^{-4}$ and the O / H$_2$O ratio is 10$^2$. The rate of dissociation per unit volume is $R_{\rm diss}$ = $G_0$ $k_{\rm diss}$ $n_{\rm H2O}$ where \mbox{$k_{\rm diss}$ = 8.0 $\times$ 10$^{-10}$ s$^{-1}$} \citep{vandishoeck06}. The water formation rate is set by the two reactions O + H$_2$ $\rightarrow$ OH + H and OH + H$_2$ $\rightarrow$ H$_2$O + H. The formation rate of water is:
\begin{equation}
\frac{{\rm d}n_{\rm H2O}}{{\rm d} t} = k_{\rm H2O}\, n_{\rm H2}\, n_{\rm OH} - G_0\, k_{\rm diss}\, n_{\rm H2O} = 0
\end{equation}
assuming equilibrium. OH is assumed to be produced rapidly in reactions between O and H$_2$ and destroyed through reactions with H$_2$; photodissociation of OH is not considered here, i.e., $n_{\rm OH} \ll n_{\rm H2O}$ and $n_{\rm O}$ \citep{wampfler13}. Thus, the formation rate is:
\begin{equation}
\frac{{\rm d}n_{\rm OH}}{{\rm d} t} = k_{\rm OH}\, n_{\rm H2}\, n_{\rm O} - k_{\rm H2O}\, n_{\rm H2}\, n_{\rm OH} = 0\ .
\end{equation}
Reordering these equations gives:
\begin{equation}
G_0 = \frac{k_{\rm OH}}{k_{\rm diss}} \frac{n_{\rm O}}{n_{\rm H2O}} n_{\rm H2}\ . 
\end{equation}
For an O / H$_2$O ratio of 10$^2$ and the $k_{\rm OH}$($T$) rate coefficient from UMIST12 \citep{mcelroy13}, this expression reduces to:
\begin{align}
G_0 &= 100 \frac{3.13\times 10^{-13}~\mathrm{s}^{-1} (T/300~\mathrm{K})^{2.7}\exp(-3150~\mathrm{K}/T)}{8.0\times 10^{-10}~\mathrm{s}^{-1}} n_{\rm H2} \nonumber\\
&= 3.91\times 10^{-2} (T/300~\mathrm{K})^{2.7}\exp(-3150~\mathrm{K}/T)\ n_{\rm H2}\ .
\end{align}

The required value of $G_0$ is shown in Fig. \ref{fig:h2o_diss_form} for three different H$_2$ densities. For a temperature of 300 K and a density of 10$^6$ cm$^{-3}$, a modest value of $G_0$ of a few is all that is required to balance the H$_2$O production and destruction to reach an equilibrium abundance of 10$^{-6}$. Naturally, these values are order-of-magnitude estimates; they may be compared to the more complete calculations of \citet{bethell09} who included OH in their simple network. The results are shown in Fig. \ref{fig:h2o_diss_form} for the same initial conditions, and there is generally excellent agreement, which in this case means less than an order of magnitude difference. These simple calculations serve to illustrate that only modest values of $G_0$ are needed to maintain low equilibrium abundances of water in low-mass protostellar sources. More sophisticated shock or wind models are required to further our understanding of water in protostars, such as those by e.g., \citet{yvart16} or \citet{melnick15}. Additionally, velocity-resolved observations of the [\ion{O}{i}] transition at 63 $\mu$m with SOFIA-GREAT will further aid in constraining the water chemistry.

\section{Summary and conclusions}
\label{sect:conclusion}

Velocity-resolved CO 16--15 spectra of 24 embedded low-mass protostars using \textit{Herschel}-HIFI have been presented. Emission is detected toward all sources, as expected from \textit{Herschel}-PACS observations of the same line. The velocity-resolved line profiles are both more complex and broader than low-$J$ CO 3--2 line profiles. 

The main conclusions are the following:
\begin{enumerate}
\item The profiles typically show multiple velocity components first detected in water line profiles with HIFI, in particular a broader component with typical $FWHM$ of 20 km\,s$^{-1}$ and a narrower component with a $FWHM$~$\sim$~5~km\,s$^{-1}$ typically blue-shifted by more than 1 km\,s$^{-1}$. Furthermore, two sources show so-called extremely-high-velocity features, or ``bullets'', also seen in lower-$J$ CO transitions and water. 
\item The broader component is identified as the 300 K component seen in the PACS CO data, where the rotational temperature is measured over the range $J_{\rm up}$ = 14 -- 25. Because of its similarity to the broad H$_2$O line profiles it is likely that H$_2$O and CO 16--15 emission both originate in the same gas in protostellar systems. This gas is either located in shocks along the outflow cavity wall in a narrow layer (called cavity shocks) or the molecular disk wind. The water abundance, as inferred from CO 16--15, is $\sim$ 2$\times$10$^{-6}$ with only a very small dependence on velocity. The smooth evolution of line profiles from lower-excited CO 3--2 through CO 6--5 and 10--9 to CO 16--15 illustrates that the 300 K cavity shock component traces a warmer and separate component of the classically observed molecular outflows. 
\item The CO excitation is thermal, up to and including CO 16--15, whereas the water excitation is modeled. The 300 K component is ubiquitous; one possible fundamental mechanism is that CO cooling turns on and becomes particularly efficient when H$_2$ cooling ceases to be efficient. At this point CO takes over and gradually brings down the temperature to ambient. Model calculations are required, and in preparation, to test this hypothesis. 
\item The narrower line profile component is responsible for the hotter component seen in the CO ladder, with $T_{\rm rot}$ $\sim$~600--800~K. This component was first identified in H$_2$O line profiles. Based on the apparent velocity offsets and the association with H$_2$O, a PDR origin solely in UV-heated cavity walls is ruled out; instead, an origin in dissociative shocks, where the dissociation of the pre-shock gas is provided by UV radiation from the accreting protostar, is a more likely explanation. If so, the hot part of the CO ladder traces cooling molecular gas just prior to the onset of H$_2$ formation. Again, model calculations are under way to verify this hypothesis. 
\item The CO ladder from $J$=1--0 to $J$=49--48 thus consists of at least three distinct physical parts, all associated with the outflow. The cold $<$ 100 K component is the entrained outflow gas typically seen in low-$J$ CO lines from the ground. The warm 300 K component is either shocks or the disk wind. Finally, the hot 600--800 K component is distinct shocks. Water is only found in the warm and hot components, not the cold outflow. Ultraviolet radiation may still play a significant role on  small spatial scales in the form of dissociating envelope gas, but only on larger scales (1000 AU) is the radiation an effective heating agent, as seen in the $^{13}$CO ladder \citep{yildiz12, yildiz15}.  
\end{enumerate}
The observations presented here demonstrate the power of, and the necessity for, velocity resolution for interpreting emission from complex systems such as low-mass protostars. Furthermore, they open a window for characterizing the more energetic part of low-mass protostellar evolution. Several open questions still remain: If H$_2$O is photodissociated, is the atomic oxygen abundance consistent with the inferred H$_2$O abundance? Does H$_2$O and high-$J$ CO emission predominantly trace currently entrained outflow material, or does emission originate in a disk wind, or something else? The next step in understanding the energetics will be to find suitable tracers observable with facilities such as ALMA, where the enormous increase in spatial resolution will pinpoint exactly where these processes occur. 

\begin{acknowledgements}
We wish to thank the entire WISH and DIGIT \textit{Herschel} teams for a myriad of stimulating discussions and always entertaining team meetings. \\

LEK and JKJ acknowledge support from the European Research Council (ERC) under the European Union's Horizon 2020 research and innovation programme (grant agreement No 646908) through ERC Consolidator Grant ``S4F''. Research at the Centre for Star and Planet Formation is funded by the Danish National Research Foundation. JCM acknowledges support from the European Research Council under the European Community's Horizon 2020 framework program (2014-2020) via the ERC Consolidator grant ``From Cloud to Star Formation (CSF)'' (project number 648505). AK acknowledges support from the Polish National Science Center grant 2013/11/N/ST9/00400.\\

Astrochemistry in Leiden is supported by the Netherlands Research School for Astronomy (NOVA), by a Royal Netherlands Academy of Arts and Sciences (KNAW) professor prize, by a Spinoza grant and grant 614.001.008 from the Netherlands Organisation for Scientific Research (NWO), and by the European Community's Seventh Framework Programme FP7/2007-2013 under grant agreement 238258 (LASSIE). \\

HIFI has been designed and built by a consortium of institutes and university departments from across Europe, Canada and the United States under the leadership of SRON Netherlands Institute for Space Research, Groningen, The Netherlands and with major contributions from Germany, France and the US. Consortium members are: Canada: CSA, U.Waterloo; France: CESR, LAB, LERMA, IRAM; Germany: KOSMA, MPIfR, MPS; Ireland, NUI Maynooth; Italy: ASI, IFSI-INAF, Osservatorio Astrofisico di Arcetri-INAF; Netherlands: SRON, TUD; Poland: CAMK, CBK; Spain: Observatorio Astron{\'o}mico Nacional (IGN), Centro de Astrobiolog{\'i}a (CSIC-INTA). Sweden: Chalmers University of Technology - MC2, RSS \& GARD; Onsala Space Observatory; Swedish National Space Board, Stockholm University - Stockholm Observatory; Switzerland: ETH Zurich, FHNW; USA: Caltech, JPL, NHSC. 
\end{acknowledgements}

\bibliographystyle{aa}
\bibliography{bibliography.bib}

\begin{thebibliography}{78}
\expandafter\ifx\csname natexlab\endcsname\relax\def\natexlab#1{#1}\fi

\bibitem[{{ALMA Partnership} {et~al.}(2015){ALMA Partnership}, {Brogan},
  {P{\'e}rez}, {Hunter}, {Dent}, {Hales}, {Hills}, {Corder}, {Fomalont},
  {Vlahakis}, {Asaki}, {Barkats}, {Hirota}, {Hodge}, {Impellizzeri}, {Kneissl},
  {Liuzzo}, {Lucas}, {Marcelino}, {Matsushita}, {Nakanishi}, {Phillips},
  {Richards}, {Toledo}, {Aladro}, {Broguiere}, {Cortes}, {Cortes}, {Espada},
  {Galarza}, {Garcia-Appadoo}, {Guzman-Ramirez}, {Humphreys}, {Jung}, {Kameno},
  {Laing}, {Leon}, {Marconi}, {Mignano}, {Nikolic}, {Nyman}, {Radiszcz},
  {Remijan}, {Rod{\'o}n}, {Sawada}, {Takahashi}, {Tilanus}, {Vila Vilaro},
  {Watson}, {Wiklind}, {Akiyama}, {Chapillon}, {de Gregorio-Monsalvo}, {Di
  Francesco}, {Gueth}, {Kawamura}, {Lee}, {Nguyen Luong}, {Mangum}, {Pietu},
  {Sanhueza}, {Saigo}, {Takakuwa}, {Ubach}, {van Kempen}, {Wootten},
  {Castro-Carrizo}, {Francke}, {Gallardo}, {Garcia}, {Gonzalez}, {Hill},
  {Kaminski}, {Kurono}, {Liu}, {Lopez}, {Morales}, {Plarre}, {Schieven},
  {Testi}, {Videla}, {Villard}, {Andreani}, {Hibbard}, \& {Tatematsu}}]{alma15}
{ALMA Partnership}, {Brogan}, C.~L., {P{\'e}rez}, L.~M., {et~al.} 2015, \apjl,
  808, L3

\bibitem[{{Andr{\'e}} {et~al.}(1990){Andr{\'e}}, {Montmerle}, {Feigelson}, \&
  {Steppe}}]{andre90}
{Andr{\'e}}, P., {Montmerle}, T., {Feigelson}, E.~D., \& {Steppe}, H. 1990,
  \aap, 240, 321

\bibitem[{{Benz} {et~al.}(2016){Benz}, {Bruderer}, {van Dishoeck}, {Melchior},
  {Wampfler}, {van der Tak}, {Goicoechea}, {Indriolo}, {Kristensen}, {Lis},
  {Mottram}, {Bergin}, {Caselli}, {Herpin}, {Hogerheijde}, {Johnstone},
  {Liseau}, {Nisini}, {Tafalla}, {Visser}, \& {Wyrowski}}]{benz16}
{Benz}, A.~O., {Bruderer}, S., {van Dishoeck}, E.~F., {et~al.} 2016, \aap, 590,
  A105

\bibitem[{{Bergin} {et~al.}(1998){Bergin}, {Melnick}, \& {Neufeld}}]{bergin98}
{Bergin}, E.~A., {Melnick}, G.~J., \& {Neufeld}, D.~A. 1998, \apj, 499, 777

\bibitem[{{Bethell} \& {Bergin}(2009)}]{bethell09}
{Bethell}, T. \& {Bergin}, E. 2009, Science, 326, 1675

\bibitem[{{Bjerkeli} {et~al.}(2012){Bjerkeli}, {Liseau}, {Larsson}, {Rydbeck},
  {Nisini}, {Tafalla}, {Antoniucci}, {Benedettini}, {Bergman}, {Cabrit},
  {Giannini}, {Melnick}, {Neufeld}, {Santangelo}, \& {van
  Dishoeck}}]{bjerkeli12}
{Bjerkeli}, P., {Liseau}, R., {Larsson}, B., {et~al.} 2012, \aap, 546, A29

\bibitem[{{Bruderer} {et~al.}(2012){Bruderer}, {van Dishoeck}, {Doty}, \&
  {Herczeg}}]{bruderer12}
{Bruderer}, S., {van Dishoeck}, E.~F., {Doty}, S.~D., \& {Herczeg}, G.~J. 2012,
  \aap, 541, A91

\bibitem[{{Cernicharo} {et~al.}(1989){Cernicharo}, {Guelin}, {Penalver},
  {Martin-Pintado}, \& {Mauersberger}}]{cernicharo89}
{Cernicharo}, J., {Guelin}, M., {Penalver}, J., {Martin-Pintado}, J., \&
  {Mauersberger}, R. 1989, \aap, 222, L1

\bibitem[{{Choi} {et~al.}(1993){Choi}, {Evans}, \& {Jaffe}}]{choi93}
{Choi}, M., {Evans}, II, N.~J., \& {Jaffe}, D.~T. 1993, \apj, 417, 624

\bibitem[{{Daniel} {et~al.}(2011){Daniel}, {Dubernet}, \&
  {Grosjean}}]{daniel11}
{Daniel}, F., {Dubernet}, M.-L., \& {Grosjean}, A. 2011, \aap, 536, A76

\bibitem[{{de Graauw} {et~al.}(2010){de Graauw}, {Helmich}, {Phillips},
  {Stutzki}, {Caux}, {Whyborn}, {Dieleman}, {Roelfsema}, {Aarts}, {Assendorp},
  {Bachiller}, {Baechtold}, {Barcia}, {Beintema}, {Belitsky}, {Benz}, {Bieber},
  {Boogert}, {Borys}, {Bumble}, {Ca{\"i}s}, {Caris}, {Cerulli-Irelli},
  {Chattopadhyay}, {Cherednichenko}, {Ciechanowicz}, {Coeur-Joly}, {Comito},
  {Cros}, {de Jonge}, {de Lange}, {Delforges}, {Delorme}, {den Boggende},
  {Desbat}, {Diez-Gonz{\'a}lez}, {di Giorgio}, {Dubbeldam}, {Edwards},
  {Eggens}, {Erickson}, {Evers}, {Fich}, {Finn}, {Franke}, {Gaier}, {Gal},
  {Gao}, {Gallego}, {Gauffre}, {Gill}, {Glenz}, {Golstein}, {Goulooze},
  {Gunsing}, {G{\"u}sten}, {Hartogh}, {Hatch}, {Higgins}, {Honingh}, {Huisman},
  {Jackson}, {Jacobs}, {Jacobs}, {Jarchow}, {Javadi}, {Jellema}, {Justen},
  {Karpov}, {Kasemann}, {Kawamura}, {Keizer}, {Kester}, {Klapwijk}, {Klein},
  {Kollberg}, {Kooi}, {Kooiman}, {Kopf}, {Krause}, {Krieg}, {Kramer},
  {Kruizenga}, {Kuhn}, {Laauwen}, {Lai}, {Larsson}, {Leduc}, {Leinz}, {Lin},
  {Liseau}, {Liu}, {Loose}, {L{\'o}pez-Fernandez}, {Lord}, {Luinge}, {Marston},
  {Mart{\'{\i}}n-Pintado}, {Maestrini}, {Maiwald}, {McCoey}, {Mehdi}, {Megej},
  {Melchior}, {Meinsma}, {Merkel}, {Michalska}, {Monstein}, {Moratschke},
  {Morris}, {Muller}, {Murphy}, {Naber}, {Natale}, {Nowosielski}, {Nuzzolo},
  {Olberg}, {Olbrich}, {Orfei}, {Orleanski}, {Ossenkopf}, {Peacock}, {Pearson},
  {Peron}, {Phillip-May}, {Piazzo}, {Planesas}, {Rataj}, {Ravera}, {Risacher},
  {Salez}, {Samoska}, {Saraceno}, {Schieder}, {Schlecht}, {Schl{\"o}der},
  {Schm{\"u}lling}, {Schultz}, {Schuster}, {Siebertz}, {Smit}, {Szczerba},
  {Shipman}, {Steinmetz}, {Stern}, {Stokroos}, {Teipen}, {Teyssier}, {Tils},
  {Trappe}, {van Baaren}, {van Leeuwen}, {van de Stadt}, {Visser}, {Wildeman},
  {Wafelbakker}, {Ward}, {Wesselius}, {Wild}, {Wulff}, {Wunsch}, {Tielens},
  {Zaal}, {Zirath}, {Zmuidzinas}, \& {Zwart}}]{degraauw10}
{de Graauw}, T., {Helmich}, F.~P., {Phillips}, T.~G., {et~al.} 2010, \aap, 518,
  L6

\bibitem[{{Dionatos} {et~al.}(2013){Dionatos}, {J{\o}rgensen}, {Green},
  {Herczeg}, {Evans}, {Kristensen}, {Lindberg}, \& {van Dishoeck}}]{dionatos13}
{Dionatos}, O., {J{\o}rgensen}, J.~K., {Green}, J.~D., {et~al.} 2013, \aap,
  558, A88

\bibitem[{{Edwards} {et~al.}(1987){Edwards}, {Cabrit}, {Strom}, {Heyer},
  {Strom}, \& {Anderson}}]{edwards87}
{Edwards}, S., {Cabrit}, S., {Strom}, S.~E., {et~al.} 1987, \apj, 321, 473

\bibitem[{{Enoch} {et~al.}(2009){Enoch}, {Evans}, {Sargent}, \&
  {Glenn}}]{enoch09}
{Enoch}, M.~L., {Evans}, N.~J., {Sargent}, A.~I., \& {Glenn}, J. 2009, \apj,
  692, 973

\bibitem[{{Fedele} {et~al.}(2013){Fedele}, {Bruderer}, {van Dishoeck},
  {Hogerheijde}, {Panic}, {Brown}, \& {Henning}}]{fedele13}
{Fedele}, D., {Bruderer}, S., {van Dishoeck}, E.~F., {et~al.} 2013, \apjl, 776,
  L3

\bibitem[{{Fillion} {et~al.}(2001){Fillion}, {van Harrevelt}, {Ruiz},
  {Castillejo}, {Zanganeh}, {Lemaire}, {van Hemert}, \& {Rostas}}]{fillion01}
{Fillion}, J.~H., {van Harrevelt}, R., {Ruiz}, J., {et~al.} 2001, The Journal
  of Physical Chemistry A, 105, 11414

\bibitem[{{Flower} \& {Pineau des For{\^e}ts}(2010)}]{flower10}
{Flower}, D.~R. \& {Pineau des For{\^e}ts}, G. 2010, \mnras, 406, 1745

\bibitem[{{Flower} \& {Pineau des For{\^e}ts}(2015)}]{flower15}
{Flower}, D.~R. \& {Pineau des For{\^e}ts}, G. 2015, \aap, 578, A63

\bibitem[{{Franklin} {et~al.}(2008){Franklin}, {Snell}, {Kaufman}, {Melnick},
  {Neufeld}, {Hollenbach}, \& {Bergin}}]{franklin08}
{Franklin}, J., {Snell}, R.~L., {Kaufman}, M.~J., {et~al.} 2008, \apj, 674,
  1015

\bibitem[{{Giovanardi} {et~al.}(1992){Giovanardi}, {Lizano}, {Natta}, {Evans},
  \& {Heiles}}]{giovanardi92}
{Giovanardi}, C., {Lizano}, S., {Natta}, A., {Evans}, II, N.~J., \& {Heiles},
  C. 1992, \apj, 397, 214

\bibitem[{{Goicoechea} {et~al.}(2012){Goicoechea}, {Cernicharo}, {Karska},
  {Herczeg}, {Polehampton}, {Wampfler}, {Kristensen}, {van Dishoeck},
  {Etxaluze}, {Bern{\'e}}, \& {Visser}}]{goicoechea12}
{Goicoechea}, J.~R., {Cernicharo}, J., {Karska}, A., {et~al.} 2012, \aap, 548,
  A77

\bibitem[{{Green} {et~al.}(2013){Green}, {Evans}, {J{\o}rgensen}, {Herczeg},
  {Kristensen}, {Lee}, {Dionatos}, {Yildiz}, {Salyk}, {Meeus}, {Bouwman},
  {Visser}, {Bergin}, {van Dishoeck}, {Rascati}, {Karska}, {van Kempen},
  {Dunham}, {Lindberg}, {Fedele}, \& {DIGIT Team1}}]{green13}
{Green}, J.~D., {Evans}, II, N.~J., {J{\o}rgensen}, J.~K., {et~al.} 2013, \apj,
  770, 123

\bibitem[{{Green} {et~al.}(2016){Green}, {Yang}, {Evans}, {Karska}, {Herczeg},
  {van Dishoeck}, {Lee}, {Larson}, \& {Bouwman}}]{green16}
{Green}, J.~D., {Yang}, Y.-L., {Evans}, II, N.~J., {et~al.} 2016, \aj, 151, 75

\bibitem[{{Harsono} {et~al.}(2013){Harsono}, {Visser}, {Bruderer}, {van
  Dishoeck}, \& {Kristensen}}]{harsono13}
{Harsono}, D., {Visser}, R., {Bruderer}, S., {van Dishoeck}, E.~F., \&
  {Kristensen}, L.~E. 2013, \aap, 555, A45

\bibitem[{{Herczeg} {et~al.}(2012){Herczeg}, {Karska}, {Bruderer},
  {Kristensen}, {van Dishoeck}, {J{\o}rgensen}, {Visser}, {Wampfler}, {Bergin},
  {Y{\i}ld{\i}z}, {Pontoppidan}, \& {Gracia-Carpio}}]{herczeg12}
{Herczeg}, G.~J., {Karska}, A., {Bruderer}, S., {et~al.} 2012, \aap, 540, A84

\bibitem[{{Hull} {et~al.}(2016){Hull}, {Girart}, {Kristensen}, {Dunham},
  {Rodr{\'{\i}}guez-Kamenetzky}, {Carrasco-Gonz{\'a}lez}, {Cort{\'e}s}, {Li},
  \& {Plambeck}}]{hull16}
{Hull}, C.~L.~H., {Girart}, J.~M., {Kristensen}, L.~E., {et~al.} 2016, \apjl,
  823, L27

\bibitem[{{Karska} {et~al.}(2013){Karska}, {Herczeg}, {van Dishoeck},
  {Wampfler}, {Kristensen}, {Goicoechea}, {Visser}, {Nisini}, {San
  Jos{\'e}-Garc{\'{\i}}a}, {Bruderer}, {{\'S}niady}, {Doty}, {Fedele},
  {Y{\i}ld{\i}z}, {Benz}, {Bergin}, {Caselli}, {Herpin}, {Hogerheijde},
  {Johnstone}, {J{\o}rgensen}, {Liseau}, {Tafalla}, {van der Tak}, \&
  {Wyrowski}}]{karska13}
{Karska}, A., {Herczeg}, G.~J., {van Dishoeck}, E.~F., {et~al.} 2013, \aap,
  552, A141

\bibitem[{{Karska} {et~al.}(2014){Karska}, {Kristensen}, {van Dishoeck},
  {Drozdovskaya}, {Mottram}, {Herczeg}, {Bruderer}, {Cabrit}, {Evans},
  {Fedele}, {Gusdorf}, {J{\o}rgensen}, {Kaufman}, {Melnick}, {Neufeld},
  {Nisini}, {Santangelo}, {Tafalla}, \& {Wampfler}}]{karska14}
{Karska}, A., {Kristensen}, L.~E., {van Dishoeck}, E.~F., {et~al.} 2014, \aap,
  572, A9

\bibitem[{{Kristensen} {et~al.}(2016){Kristensen}, {Brown}, {Wilner}, \&
  {Salyk}}]{kristensen16}
{Kristensen}, L.~E., {Brown}, J.~M., {Wilner}, D., \& {Salyk}, C. 2016, \apjl,
  822, L20

\bibitem[{{Kristensen} {et~al.}(2007){Kristensen}, {Ravkilde}, {Field},
  {Lemaire}, \& {Pineau des For{\^e}ts}}]{kristensen07}
{Kristensen}, L.~E., {Ravkilde}, T.~L., {Field}, D., {Lemaire}, J.~L., \&
  {Pineau des For{\^e}ts}, G. 2007, \aap, 469, 561

\bibitem[{{Kristensen} {et~al.}(2013){Kristensen}, {van Dishoeck}, {Benz},
  {Bruderer}, {Visser}, \& {Wampfler}}]{kristensen13}
{Kristensen}, L.~E., {van Dishoeck}, E.~F., {Benz}, A.~O., {et~al.} 2013, \aap,
  557, A23

\bibitem[{{Kristensen} {et~al.}(2012){Kristensen}, {van Dishoeck}, {Bergin},
  {Visser}, {Y{\i}ld{\i}z}, {San Jose-Garcia}, {J{\o}rgensen}, {Herczeg},
  {Johnstone}, {Wampfler}, {Benz}, {Bruderer}, {Cabrit}, {Caselli}, {Doty},
  {Harsono}, {Herpin}, {Hogerheijde}, {Karska}, {van Kempen}, {Liseau},
  {Nisini}, {Tafalla}, {van der Tak}, \& {Wyrowski}}]{kristensen12}
{Kristensen}, L.~E., {van Dishoeck}, E.~F., {Bergin}, E.~A., {et~al.} 2012,
  \aap, 542, A8

\bibitem[{{Kristensen} {et~al.}(2010){Kristensen}, {Visser}, {van Dishoeck},
  {Y{\i}ld{\i}z}, {Doty}, {Herczeg}, {Liu}, {Parise}, {J{\o}rgensen}, {van
  Kempen}, {Brinch}, {Wampfler}, {Bruderer}, {Benz}, {Hogerheijde}, {Deul},
  {Bachiller}, {Baudry}, {Benedettini}, {Bergin}, {Bjerkeli}, {Blake},
  {Bontemps}, {Braine}, {Caselli}, {Cernicharo}, {Codella}, {Daniel}, {de
  Graauw}, {di Giorgio}, {Dominik}, {Encrenaz}, {Fich}, {Fuente}, {Giannini},
  {Goicoechea}, {Helmich}, {Herpin}, {Jacq}, {Johnstone}, {Kaufman}, {Larsson},
  {Lis}, {Liseau}, {Marseille}, {McCoey}, {Melnick}, {Neufeld}, {Nisini},
  {Olberg}, {Pearson}, {Plume}, {Risacher}, {Santiago-Garc{\'{\i}}a},
  {Saraceno}, {Shipman}, {Tafalla}, {Tielens}, {van der Tak}, {Wyrowski},
  {Beintema}, {de Jonge}, {Dieleman}, {Ossenkopf}, {Roelfsema}, {Stutzki}, \&
  {Whyborn}}]{kristensen10}
{Kristensen}, L.~E., {Visser}, R., {van Dishoeck}, E.~F., {et~al.} 2010, \aap,
  521, L30

\bibitem[{{Lee} {et~al.}(2014){Lee}, {Lee}, {Bergin}, \& {Park}}]{lee14}
{Lee}, S., {Lee}, J.-E., {Bergin}, E.~A., \& {Park}, Y.-S. 2014, \apjs, 213, 33

\bibitem[{{Lefloch} {et~al.}(2012){Lefloch}, {Cabrit}, {Busquet}, {Codella},
  {Ceccarelli}, {Cernicharo}, {Pardo}, {Benedettini}, {Lis}, \&
  {Nisini}}]{lefloch12}
{Lefloch}, B., {Cabrit}, S., {Busquet}, G., {et~al.} 2012, \apjl, 757, L25

\bibitem[{{Lindberg} \& {J{\o}rgensen}(2012)}]{lindberg12}
{Lindberg}, J.~E. \& {J{\o}rgensen}, J.~K. 2012, \aap, 548, A24

\bibitem[{{Lizano} \& {Giovanardi}(1995)}]{lizano95}
{Lizano}, S. \& {Giovanardi}, C. 1995, \apj, 447, 742

\bibitem[{{Lizano} {et~al.}(1988){Lizano}, {Heiles}, {Rodriguez}, {Koo}, {Shu},
  {Hasegawa}, {Hayashi}, \& {Mirabel}}]{lizano88}
{Lizano}, S., {Heiles}, C., {Rodriguez}, L.~F., {et~al.} 1988, \apj, 328, 763

\bibitem[{{Manoj} {et~al.}(2016){Manoj}, {Green}, {Megeath}, {Evans}, {Stutz},
  {Tobin}, {Watson}, {Fischer}, {Furlan}, \& {Henning}}]{manoj16}
{Manoj}, P., {Green}, J.~D., {Megeath}, S.~T., {et~al.} 2016, \apj, 831, 69

\bibitem[{{Manoj} {et~al.}(2013){Manoj}, {Watson}, {Neufeld}, {Megeath},
  {Vavrek}, {Yu}, {Visser}, {Bergin}, {Fischer}, {Tobin}, {Stutz}, {Ali},
  {Wilson}, {Di Francesco}, {Osorio}, {Maret}, \& {Poteet}}]{manoj13}
{Manoj}, P., {Watson}, D.~M., {Neufeld}, D.~A., {et~al.} 2013, \apj, 763, 83

\bibitem[{{Maret} {et~al.}(2009){Maret}, {Bergin}, {Neufeld}, {Green},
  {Watson}, {Harwit}, {Kristensen}, {Melnick}, {Sonnentrucker}, {Tolls},
  {Werner}, {Willacy}, \& {Yuan}}]{maret09}
{Maret}, S., {Bergin}, E.~A., {Neufeld}, D.~A., {et~al.} 2009, \apj, 698, 1244

\bibitem[{{Matuszak} {et~al.}(2015){Matuszak}, {Karska}, {Kristensen},
  {Herczeg}, {Tychoniec}, {van Kempen}, \& {Fuente}}]{matuszak15}
{Matuszak}, M., {Karska}, A., {Kristensen}, L.~E., {et~al.} 2015, \aap, 578,
  A20

\bibitem[{{McElroy} {et~al.}(2013){McElroy}, {Walsh}, {Markwick}, {Cordiner},
  {Smith}, \& {Millar}}]{mcelroy13}
{McElroy}, D., {Walsh}, C., {Markwick}, A.~J., {et~al.} 2013, \aap, 550, A36

\bibitem[{{Melnick} \& {Kaufman}(2015)}]{melnick15}
{Melnick}, G.~J. \& {Kaufman}, M.~J. 2015, \apj, 806, 227

\bibitem[{{Mottram} {et~al.}(2014){Mottram}, {Kristensen}, {van Dishoeck},
  {Bruderer}, {San Jos{\'e}-Garc{\'{\i}}a}, {Karska}, {Visser}, {Santangelo},
  {Benz}, {Bergin}, {Caselli}, {Herpin}, {Hogerheijde}, {Johnstone}, {van
  Kempen}, {Liseau}, {Nisini}, {Tafalla}, {van der Tak}, \&
  {Wyrowski}}]{mottram14}
{Mottram}, J.~C., {Kristensen}, L.~E., {van Dishoeck}, E.~F., {et~al.} 2014,
  \aap, 572, A21

\bibitem[{{Neufeld}(2010)}]{neufeld10}
{Neufeld}, D.~A. 2010, \apj, 708, 635

\bibitem[{{Neufeld}(2012)}]{neufeld12}
{Neufeld}, D.~A. 2012, \apj, 749, 125

\bibitem[{{Neufeld} \& {Dalgarno}(1989)}]{neufeld89}
{Neufeld}, D.~A. \& {Dalgarno}, A. 1989, \apj, 340, 869

\bibitem[{{Neufeld} \& {Kaufman}(1993)}]{neufeld93}
{Neufeld}, D.~A. \& {Kaufman}, M.~J. 1993, \apj, 418, 263

\bibitem[{{Nisini} {et~al.}(2010){Nisini}, {Benedettini}, {Codella},
  {Giannini}, {Liseau}, {Neufeld}, {Tafalla}, {van Dishoeck}, {Bachiller},
  {Baudry}, {Benz}, {Bergin}, {Bjerkeli}, {Blake}, {Bontemps}, {Braine},
  {Bruderer}, {Caselli}, {Cernicharo}, {Daniel}, {Encrenaz}, {di Giorgio},
  {Dominik}, {Doty}, {Fich}, {Fuente}, {Goicoechea}, {de Graauw}, {Helmich},
  {Herczeg}, {Herpin}, {Hogerheijde}, {Jacq}, {Johnstone}, {J{\o}rgensen},
  {Kaufman}, {Kristensen}, {Larsson}, {Lis}, {Marseille}, {McCoey}, {Melnick},
  {Olberg}, {Parise}, {Pearson}, {Plume}, {Risacher}, {Santiago}, {Saraceno},
  {Shipman}, {van Kempen}, {Visser}, {Viti}, {Wampfler}, {Wyrowski}, {van der
  Tak}, {Y{\i}ld{\i}z}, {Delforge}, {Desbat}, {Hatch}, {P{\'e}ron}, {Schieder},
  {Stern}, {Teyssier}, \& {Whyborn}}]{nisini10}
{Nisini}, B., {Benedettini}, M., {Codella}, C., {et~al.} 2010, \aap, 518, L120

\bibitem[{{Nisini} {et~al.}(2015){Nisini}, {Santangelo}, {Giannini},
  {Antoniucci}, {Cabrit}, {Codella}, {Davis}, {Eisl{\"o}ffel}, {Kristensen},
  {Herczeg}, {Neufeld}, \& {van Dishoeck}}]{nisini15}
{Nisini}, B., {Santangelo}, G., {Giannini}, T., {et~al.} 2015, \apj, 801, 121

\bibitem[{{Ott}(2010)}]{ott10}
{Ott}, S. 2010, in Astronomical Society of the Pacific Conference Series, Vol.
  434, Astronomical Data Analysis Software and Systems XIX, ed. {Y.~Mizumoto,
  K.-I.~Morita, \& M.~Ohishi}, 139

\bibitem[{{Panoglou} {et~al.}(2012){Panoglou}, {Cabrit}, {Pineau des
  For{\^e}ts}, {Garcia}, {Ferreira}, \& {Casse}}]{panoglou12}
{Panoglou}, D., {Cabrit}, S., {Pineau des For{\^e}ts}, G., {et~al.} 2012, \aap,
  538, A2

\bibitem[{{Pilbratt} {et~al.}(2010){Pilbratt}, {Riedinger}, {Passvogel},
  {Crone}, {Doyle}, {Gageur}, {Heras}, {Jewell}, {Metcalfe}, {Ott}, \&
  {Schmidt}}]{pilbratt10}
{Pilbratt}, G.~L., {Riedinger}, J.~R., {Passvogel}, T., {et~al.} 2010, \aap,
  518, L1

\bibitem[{{Rodriguez} {et~al.}(1990){Rodriguez}, {Escalante}, {Lizano},
  {Canto}, \& {Mirabel}}]{rodriguez90}
{Rodriguez}, L.~F., {Escalante}, V., {Lizano}, S., {Canto}, J., \& {Mirabel},
  I.~F. 1990, \apj, 365, 261

\bibitem[{{Roelfsema} {et~al.}(2012){Roelfsema}, {Helmich}, {Teyssier},
  {Ossenkopf}, {Morris}, {Olberg}, {Shipman}, {Risacher}, {Akyilmaz},
  {Assendorp}, {Avruch}, {Beintema}, {Biver}, {Boogert}, {Borys}, {Braine},
  {Caris}, {Caux}, {Cernicharo}, {Coeur-Joly}, {Comito}, {de Lange},
  {Delforge}, {Dieleman}, {Dubbeldam}, {de Graauw}, {Edwards}, {Fich},
  {Flederus}, {Gal}, {di Giorgio}, {Herpin}, {Higgins}, {Hoac}, {Huisman},
  {Jarchow}, {Jellema}, {de Jonge}, {Kester}, {Klein}, {Kooi}, {Kramer},
  {Laauwen}, {Larsson}, {Leinz}, {Lord}, {Lorenzani}, {Luinge}, {Marston},
  {Mart{\'{\i}}n-Pintado}, {McCoey}, {Melchior}, {Michalska}, {Moreno},
  {M{\"u}ller}, {Nowosielski}, {Okada}, {Orlea{\'n}ski}, {Phillips}, {Pearson},
  {Rabois}, {Ravera}, {Rector}, {Rengel}, {Sagawa}, {Salomons},
  {S{\'a}nchez-Su{\'a}rez}, {Schieder}, {Schl{\"o}der}, {Schm{\"u}lling},
  {Soldati}, {Stutzki}, {Thomas}, {Tielens}, {Vastel}, {Wildeman}, {Xie},
  {Xilouris}, {Wafelbakker}, {Whyborn}, {Zaal}, {Bell}, {Bjerkeli}, {De Beck},
  {Cavali{\'e}}, {Crockett}, {Hily-Blant}, {Kama}, {Kaminski}, {Lefl{\'o}ch},
  {Lombaert}, {de Luca}, {Makai}, {Marseille}, {Nagy}, {Pacheco}, {van der
  Wiel}, {Wang}, \& {Y{\i}ld{\i}z}}]{roelfsema12}
{Roelfsema}, P.~R., {Helmich}, F.~P., {Teyssier}, D., {et~al.} 2012, \aap, 537,
  A17

\bibitem[{{Rudolph}(1992)}]{rudolph92}
{Rudolph}, A. 1992, \apjl, 397, L111

\bibitem[{{San Jos{\'e}-Garc{\'{\i}}a} {et~al.}(2013){San
  Jos{\'e}-Garc{\'{\i}}a}, {Mottram}, {Kristensen}, {van Dishoeck},
  {Y{\i}ld{\i}z}, {van der Tak}, {Herpin}, {Visser}, {McCoey}, {Wyrowski},
  {Braine}, \& {Johnstone}}]{sanjosegarcia13}
{San Jos{\'e}-Garc{\'{\i}}a}, I., {Mottram}, J.~C., {Kristensen}, L.~E.,
  {et~al.} 2013, \aap, 553, A125

\bibitem[{{Santangelo} {et~al.}(2013){Santangelo}, {Nisini}, {Antoniucci},
  {Codella}, {Cabrit}, {Giannini}, {Herczeg}, {Liseau}, {Tafalla}, \& {van
  Dishoeck}}]{santangelo13}
{Santangelo}, G., {Nisini}, B., {Antoniucci}, S., {et~al.} 2013, \aap, 557, A22

\bibitem[{{Santangelo} {et~al.}(2012){Santangelo}, {Nisini}, {Giannini},
  {Antoniucci}, {Vasta}, {Codella}, {Lorenzani}, {Tafalla}, {Liseau}, {van
  Dishoeck}, \& {Kristensen}}]{santangelo12}
{Santangelo}, G., {Nisini}, B., {Giannini}, T., {et~al.} 2012, \aap, 538, A45

\bibitem[{{Sch{\"o}ier} {et~al.}(2005){Sch{\"o}ier}, {van der Tak}, {van
  Dishoeck}, \& {Black}}]{schoier05}
{Sch{\"o}ier}, F.~L., {van der Tak}, F.~F.~S., {van Dishoeck}, E.~F., \&
  {Black}, J.~H. 2005, \aap, 432, 369

\bibitem[{{Suutarinen} {et~al.}(2014){Suutarinen}, {Kristensen}, {Mottram},
  {Fraser}, \& {van Dishoeck}}]{suutarinen14}
{Suutarinen}, A.~N., {Kristensen}, L.~E., {Mottram}, J.~C., {Fraser}, H.~J., \&
  {van Dishoeck}, E.~F. 2014, \mnras, 440, 1844

\bibitem[{{Tafalla} {et~al.}(2013){Tafalla}, {Liseau}, {Nisini}, {Bachiller},
  {Santiago-Garc{\'{\i}}a}, {van Dishoeck}, {Kristensen}, {Herczeg}, \&
  {Y{\i}ld{\i}z}}]{tafalla13}
{Tafalla}, M., {Liseau}, R., {Nisini}, B., {et~al.} 2013, \aap, 551, A116

\bibitem[{{Tafalla} {et~al.}(2010){Tafalla}, {Santiago-Garc{\'{\i}}a}, {Hacar},
  \& {Bachiller}}]{tafalla10}
{Tafalla}, M., {Santiago-Garc{\'{\i}}a}, J., {Hacar}, A., \& {Bachiller}, R.
  2010, \aap, 522, A91

\bibitem[{{van der Tak} {et~al.}(2007){van der Tak}, {Black}, {Sch{\"o}ier},
  {Jansen}, \& {van Dishoeck}}]{vandertak07}
{van der Tak}, F.~F.~S., {Black}, J.~H., {Sch{\"o}ier}, F.~L., {Jansen}, D.~J.,
  \& {van Dishoeck}, E.~F. 2007, \aap, 468, 627

\bibitem[{{van Dishoeck} {et~al.}(2014){van Dishoeck}, {Bergin}, {Lis}, \&
  {Lunine}}]{vandishoeck14}
{van Dishoeck}, E.~F., {Bergin}, E.~A., {Lis}, D.~C., \& {Lunine}, J.~I. 2014,
  Protostars and Planets VI, 835

\bibitem[{{van Dishoeck} {et~al.}(2006){van Dishoeck}, {Jonkheid}, \& {van
  Hemert}}]{vandishoeck06}
{van Dishoeck}, E.~F., {Jonkheid}, B., \& {van Hemert}, M.~C. 2006, Faraday
  Discussions, 133, 231

\bibitem[{{van Dishoeck} {et~al.}(2011){van Dishoeck}, {Kristensen}, {Benz},
  {Bergin}, {Caselli}, {Cernicharo}, {Herpin}, {Hogerheijde}, {Johnstone},
  {Liseau}, {Nisini}, {Shipman}, {Tafalla}, {van der Tak}, {Wyrowski},
  {Aikawa}, {Bachiller}, {Baudry}, {Benedettini}, {Bjerkeli}, {Blake},
  {Bontemps}, {Braine}, {Brinch}, {Bruderer}, {Chavarr{\'{\i}}a}, {Codella},
  {Daniel}, {de Graauw}, {Deul}, {di Giorgio}, {Dominik}, {Doty}, {Dubernet},
  {Encrenaz}, {Feuchtgruber}, {Fich}, {Frieswijk}, {Fuente}, {Giannini},
  {Goicoechea}, {Helmich}, {Herczeg}, {Jacq}, {J{\o}rgensen}, {Karska},
  {Kaufman}, {Keto}, {Larsson}, {Lefloch}, {Lis}, {Marseille}, {McCoey},
  {Melnick}, {Neufeld}, {Olberg}, {Pagani}, {Pani{\'c}}, {Parise}, {Pearson},
  {Plume}, {Risacher}, {Salter}, {Santiago-Garc{\'{\i}}a}, {Saraceno},
  {St{\"a}uber}, {van Kempen}, {Visser}, {Viti}, {Walmsley}, {Wampfler}, \&
  {Y{\i}ld{\i}z}}]{vandishoeck11}
{van Dishoeck}, E.~F., {Kristensen}, L.~E., {Benz}, A.~O., {et~al.} 2011,
  \pasp, 123, 138

\bibitem[{{van Kempen} {et~al.}(2010){van Kempen}, {Kristensen}, {Herczeg},
  {Visser}, {van Dishoeck}, {Wampfler}, {Bruderer}, {Benz}, {Doty}, {Brinch},
  {Hogerheijde}, {J{\o}rgensen}, {Tafalla}, {Neufeld}, {Bachiller}, {Baudry},
  {Benedettini}, {Bergin}, {Bjerkeli}, {Blake}, {Bontemps}, {Braine},
  {Caselli}, {Cernicharo}, {Codella}, {Daniel}, {di Giorgio}, {Dominik},
  {Encrenaz}, {Fich}, {Fuente}, {Giannini}, {Goicoechea}, {de Graauw},
  {Helmich}, {Herpin}, {Jacq}, {Johnstone}, {Kaufman}, {Larsson}, {Lis},
  {Liseau}, {Marseille}, {McCoey}, {Melnick}, {Nisini}, {Olberg}, {Parise},
  {Pearson}, {Plume}, {Risacher}, {Santiago-Garc{\'{\i}}a}, {Saraceno},
  {Shipman}, {van der Tak}, {Wyrowski}, {Y{\i}ld{\i}z}, {Ciechanowicz},
  {Dubbeldam}, {Glenz}, {Huisman}, {Lin}, {Morris}, {Murphy}, \&
  {Trappe}}]{vankempen10}
{van Kempen}, T.~A., {Kristensen}, L.~E., {Herczeg}, G.~J., {et~al.} 2010,
  \aap, 518, L121

\bibitem[{{van Kempen} {et~al.}(2009){van Kempen}, {van Dishoeck},
  {G{\"u}sten}, {Kristensen}, {Schilke}, {Hogerheijde}, {Boland}, {Nefs},
  {Menten}, {Baryshev}, \& {Wyrowski}}]{vankempen09}
{van Kempen}, T.~A., {van Dishoeck}, E.~F., {G{\"u}sten}, R., {et~al.} 2009,
  \aap, 501, 633

\bibitem[{{Visser} {et~al.}(2012){Visser}, {Kristensen}, {Bruderer}, {van
  Dishoeck}, {Herczeg}, {Brinch}, {Doty}, {Harsono}, \& {Wolfire}}]{visser12}
{Visser}, R., {Kristensen}, L.~E., {Bruderer}, S., {et~al.} 2012, \aap, 537,
  A55

\bibitem[{{Wampfler} {et~al.}(2013){Wampfler}, {Bruderer}, {Karska}, {Herczeg},
  {van Dishoeck}, {Kristensen}, {Goicoechea}, {Benz}, {Doty}, {McCoey},
  {Baudry}, {Giannini}, \& {Larsson}}]{wampfler13}
{Wampfler}, S.~F., {Bruderer}, S., {Karska}, A., {et~al.} 2013, \aap, 552, A56

\bibitem[{{Yang} {et~al.}(2010){Yang}, {Stancil}, {Balakrishnan}, \&
  {Forrey}}]{yang10}
{Yang}, B., {Stancil}, P.~C., {Balakrishnan}, N., \& {Forrey}, R.~C. 2010,
  \apj, 718, 1062

\bibitem[{{Yang} {et~al.}(2017){Yang}, {Evans}, {Green}, {Dunham}, \&
  {J{\o}rgensen}}]{yang17}
{Yang}, Y.-L., {Evans}, II, N.~J., {Green}, J.~D., {Dunham}, M.~M., \&
  {J{\o}rgensen}, J.~K. 2017, \apj, 835, 259

\bibitem[{{Y{\i}ld{\i}z} {et~al.}(2012){Y{\i}ld{\i}z}, {Kristensen}, {van
  Dishoeck}, {Belloche}, {van Kempen}, {Hogerheijde}, {G{\"u}sten}, \& {van der
  Marel}}]{yildiz12}
{Y{\i}ld{\i}z}, U.~A., {Kristensen}, L.~E., {van Dishoeck}, E.~F., {et~al.}
  2012, \aap, 542, A86

\bibitem[{{Y{\i}ld{\i}z} {et~al.}(2015){Y{\i}ld{\i}z}, {Kristensen}, {van
  Dishoeck}, {Hogerheijde}, {Karska}, {Belloche}, {Endo}, {Frieswijk},
  {G{\"u}sten}, {van Kempen}, {Leurini}, {Nagy}, {P{\'e}rez-Beaupuits},
  {Risacher}, {van der Marel}, {van Weeren}, \& {Wyrowski}}]{yildiz15}
{Y{\i}ld{\i}z}, U.~A., {Kristensen}, L.~E., {van Dishoeck}, E.~F., {et~al.}
  2015, \aap, 576, A109

\bibitem[{{Y{\i}ld{\i}z} {et~al.}(2013){Y{\i}ld{\i}z}, {Kristensen}, {van
  Dishoeck}, {San Jos{\'e}-Garc{\'i}a}, {Karska}, {Harsono}, {Tafalla},
  {Fuente}, {Visser}, {J{\o}rgensen}, \& {Hogerheijde}}]{yildiz13}
{Y{\i}ld{\i}z}, U.~A., {Kristensen}, L.~E., {van Dishoeck}, E.~F., {et~al.}
  2013, \aap, 556, A89

\bibitem[{{Yvart} {et~al.}(2016){Yvart}, {Cabrit}, {Pineau des For{\^e}ts}, \&
  {Ferreira}}]{yvart16}
{Yvart}, W., {Cabrit}, S., {Pineau des For{\^e}ts}, G., \& {Ferreira}, J. 2016,
  \aap, 585, A74

\end{thebibliography}

\appendix

\section{Observational details}

Observing times and \textit{Herschel} obsid's are provided in Table \ref{tab:obsid}, as are the rms values as measured in 1 km\,s$^{-1}$ channels. 

\begin{figure*}[!t]
\includegraphics[width=0.95\textwidth]{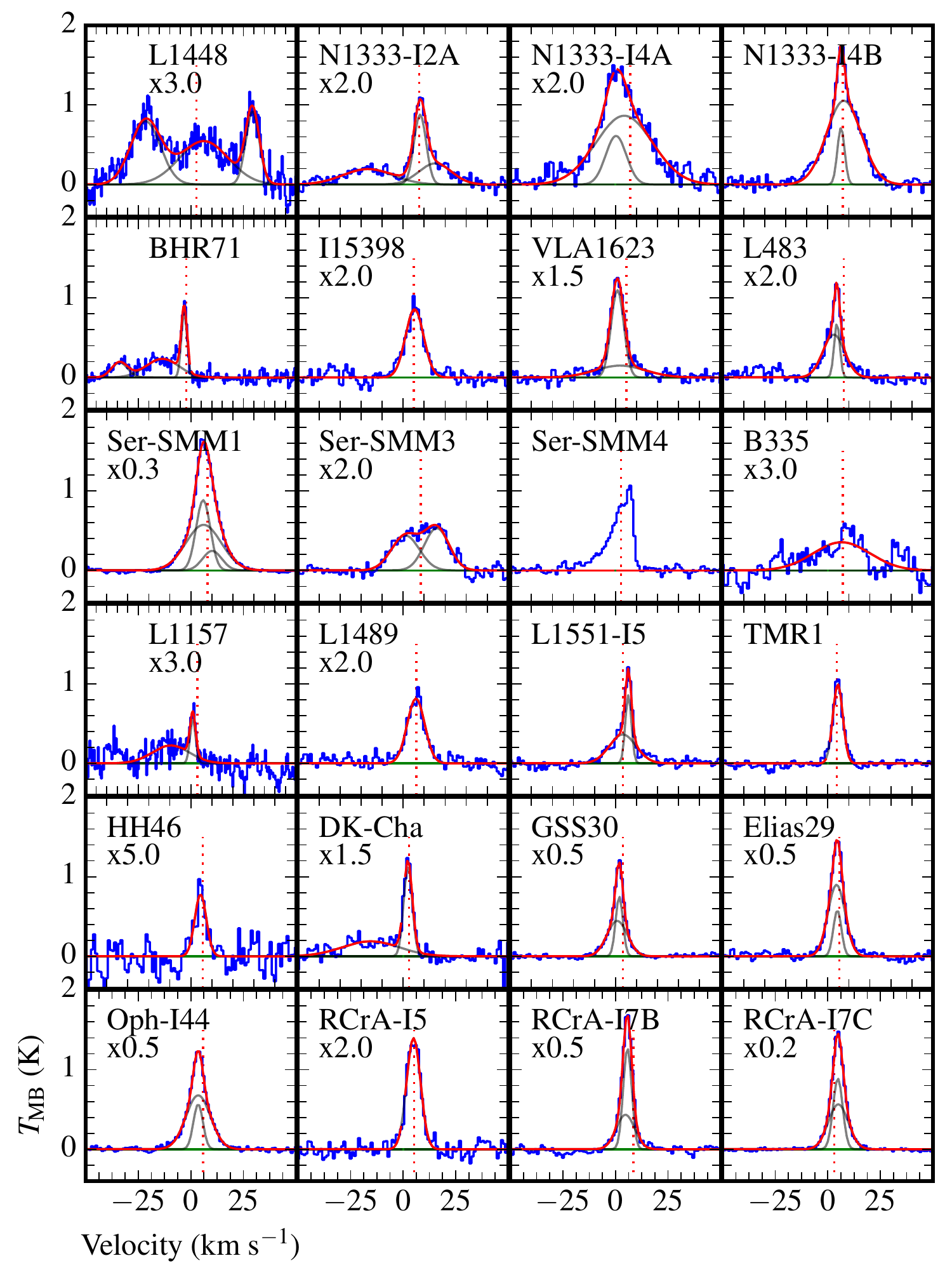}
\caption{CO 16--15 spectra toward all observed sources. The source velocity is marked with a red dashed line in each panel and the baseline is shown in green. The best-fit Gaussian decomposition is overlaid in red, with individual Gaussian components shown in gray. The velocity scale in the L1448, BHR71, and L1157 spectra is from --100 to $+$100 km\,s$^{-1}$. Ser-SMM4 is not decomposed into Gaussian functions because it appears entirely triangular. \label{fig:gauss}}
\end{figure*}

\begin{figure}
\includegraphics[width=0.95\columnwidth]{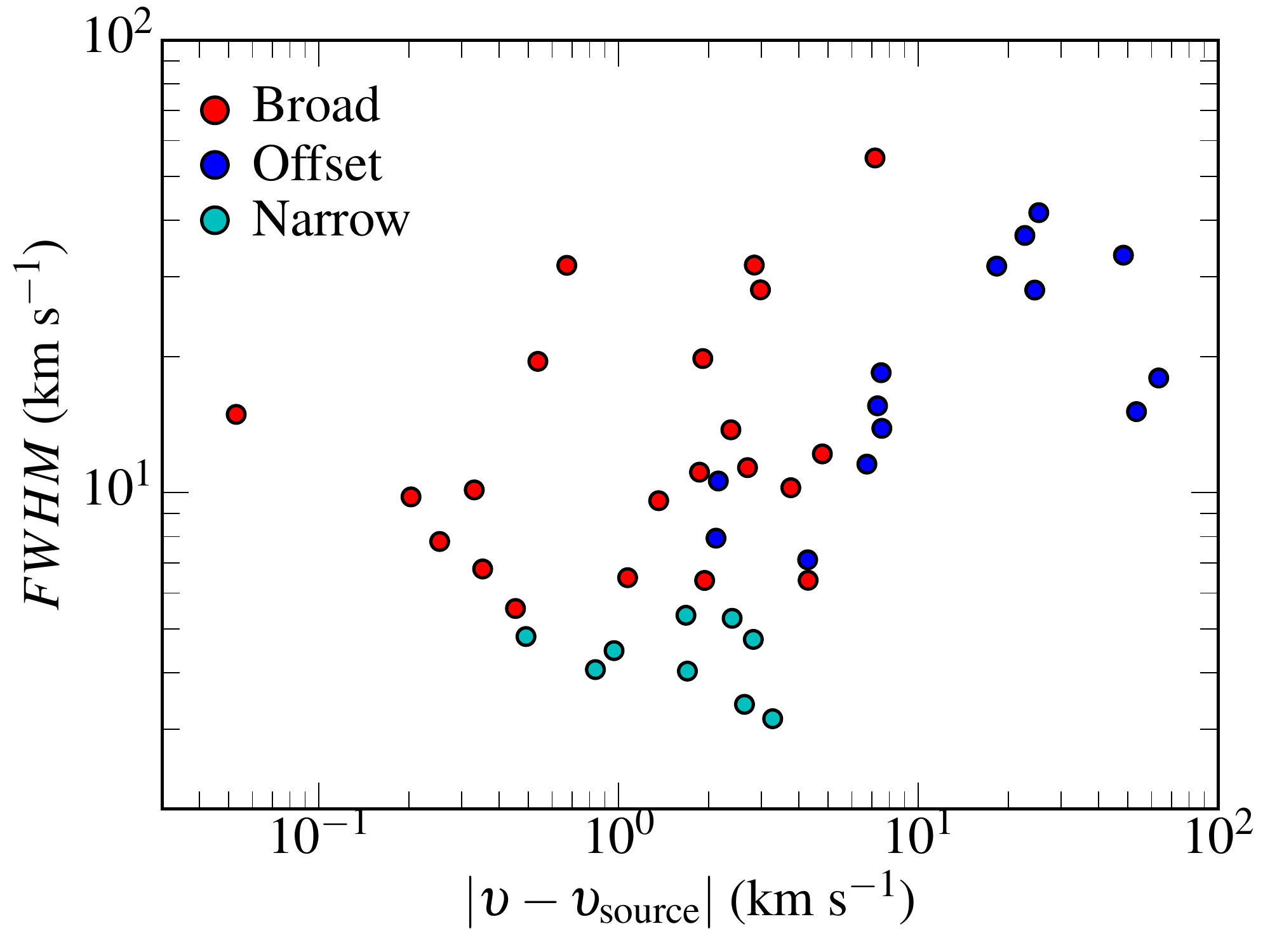}
\caption{Parameter space spanned by the various Gaussian components identified and labeled in Table \ref{tab:gauss}. \label{fig:comp_space}}
\end{figure}

\begin{figure*}[!t]
\begin{center}
\includegraphics[width=0.95\textwidth]{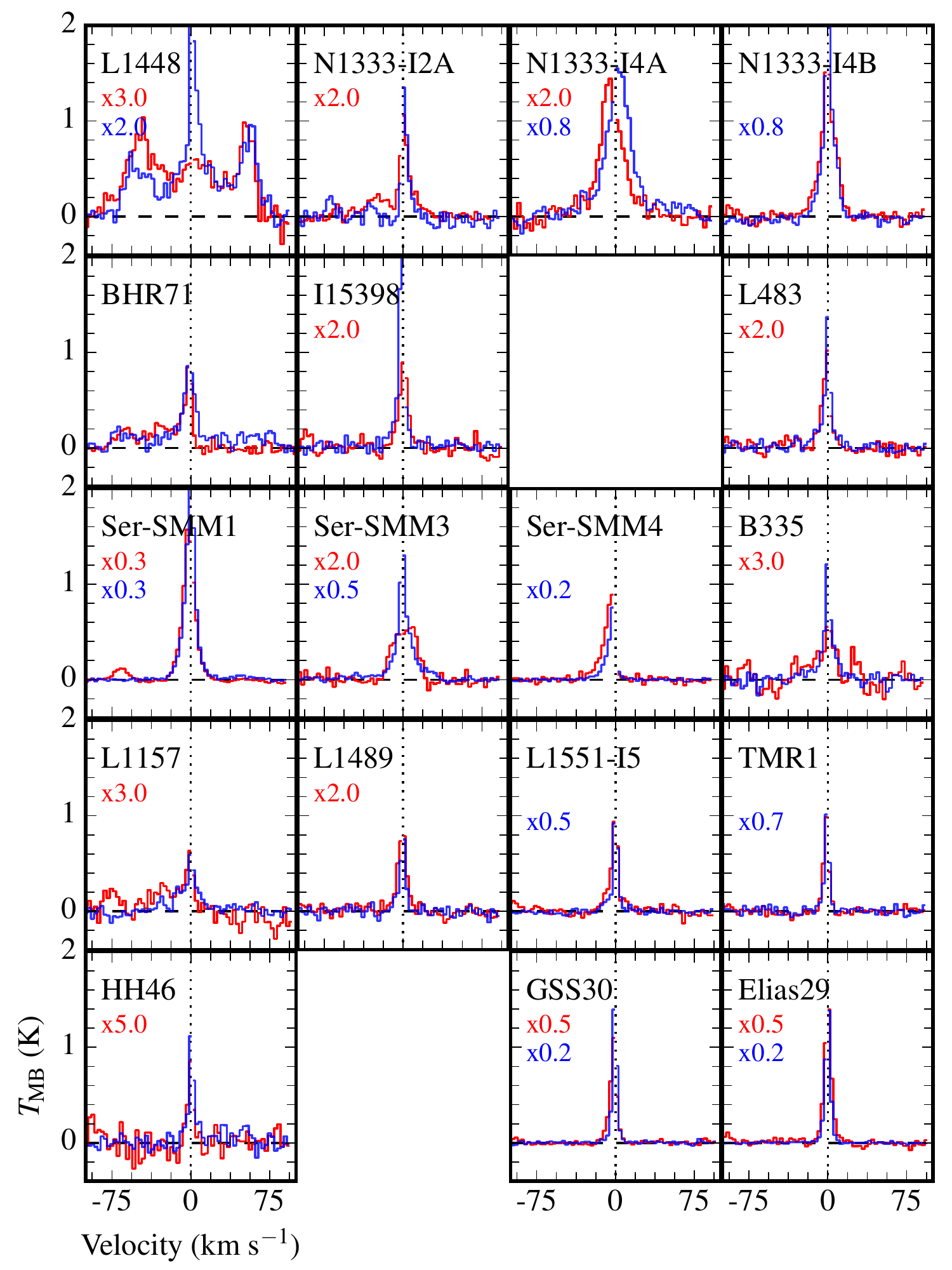}
\caption{CO 16--15 (red) and 10--9 (blue) line spectra for all 19 sources where both data sets are available shown individually. The spectra have been rebinned to 3 km\,s$^{-1}$ channels. The vertical dotted line is for 0 km\,s$^{-1}$ and the horizontal dashed line is for the baseline. Both spectra have been scaled for clarity, with the scaling factor shown in each plot.\label{fig:trot_ind}}
\end{center}
\end{figure*}

\begin{figure*}[!t]
\begin{center}
\includegraphics[width=0.95\textwidth]{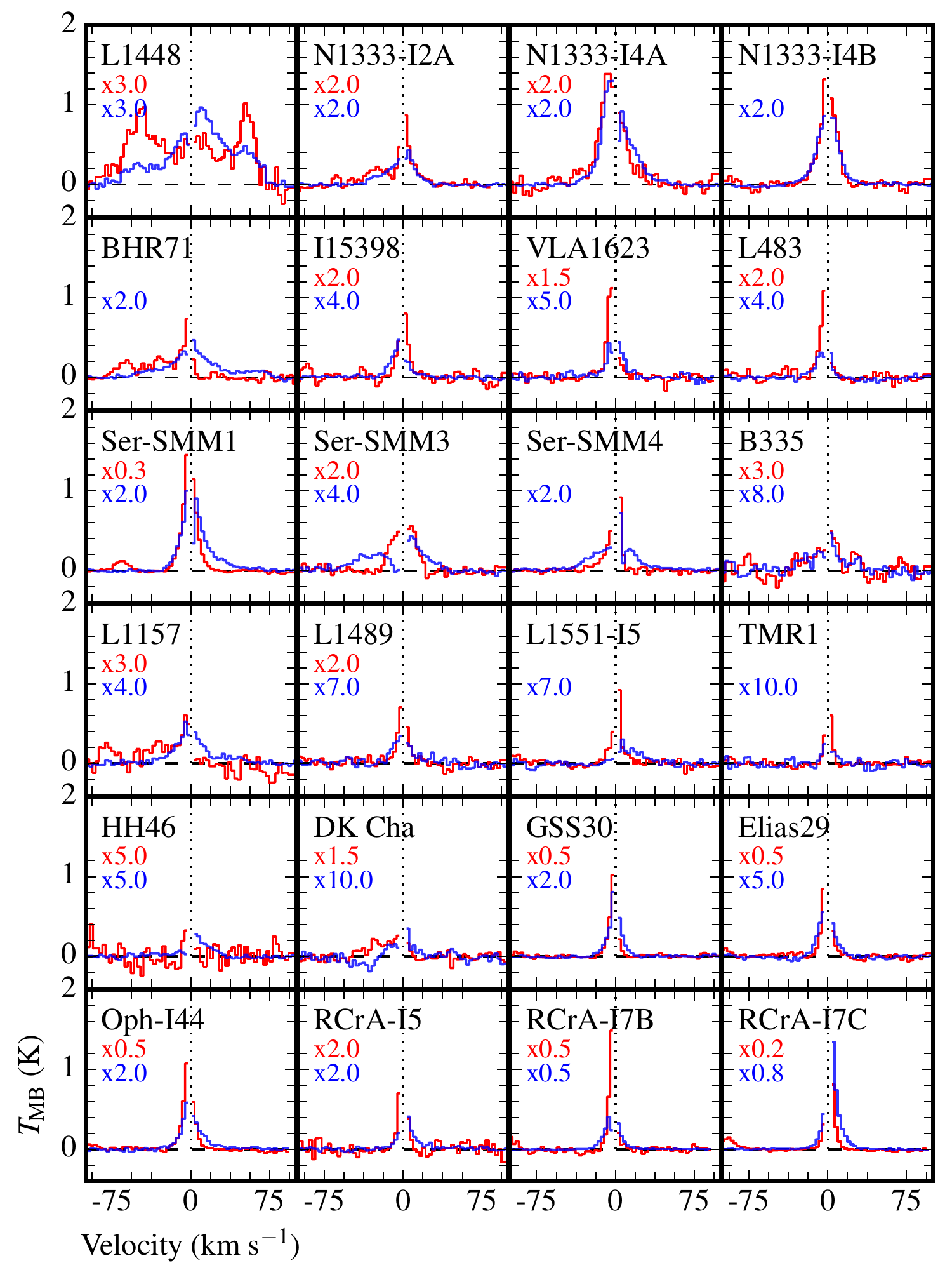}
\caption{H$_2$O 1$_{10}$--1$_{01}$ (blue) and CO 16--15 (red) line profiles for all sources shown individually. The spectra have been rebinned to 3 km\,s$^{-1}$ channels. The central 2 km\,s$^{-1}$ have been masked out because of self-absorption in the H$_2$O line, and are not shown. The vertical dotted line is for 0 km\,s$^{-1}$ and the horizontal dashed line is the baseline. Both the H$_2$O and CO spectra have been scaled for clarity, with the scaling factor shown in each plot. \label{fig:h2o_ind}}
\end{center}
\end{figure*}

Figure \ref{fig:gauss} shows the Gaussian fits to each line profile, where the fit parameters are listed in Table \ref{tab:gauss}. Figure \ref{fig:comp_space} shows the parameter space spanned by the different components. Figures \ref{fig:trot_ind} and \ref{fig:h2o_ind} show the CO 16--15 CO 10--9, and H$_2$O 1$_{10}$--1$_{01}$ line profiles overplotted on one another for each source. 

\begin{table}
\caption{Observation details. \label{tab:obsid}}
\tiny
\begin{center}
\begin{tabular}{l c c c c} \hline \hline
Source & obsid & $t_{\rm int}$ & rms & Program \\
& & (min) & (mK) & \\ \hline
L1448-MM	& 1342249001 & 31 & 57 & \verb+OT2_lkrist01_2+ \\
N1333-IRAS2A	& 1342249004 & 92 & 42 & \verb+OT2_lkrist01_2+ \\
N1333-IRAS4A	& 1342249003 & 31 & 83 & \verb+OT2_lkrist01_2+ \\
N1333-IRAS4B	& 1342249002 & 20 & 93 & \verb+OT2_lkrist01_2+ \\
BHR71		& 1342252148 & 20 & 125 & \verb+OT2_lkrist01_2+ \\
IRAS15398	& 1342249655 & 45 & 58 & \verb+OT2_lkrist01_2+ \\
VLA1623		& 1342251069 & 45 & 63 & \verb+OT2_lkrist01_2+ \\
L483			& 1342252166 & 60 & 52 & \verb+OT2_lkrist01_2+ \\
Ser-SMM1	& 1342218494 & 50 & 56 & \verb+KPGT_evandish_1+ \\
Ser-SMM4	& 1342252164 & 60 & 58 & \verb+OT2_lkrist01_2+ \\
Ser-SMM3	& 1342252165 & 60 & 48 & \verb+OT2_lkrist01_2+ \\
B335			& 1342245314 & 40 & 63 & \verb+OT2_lkrist01_2+ \\
L1157		& 1342246025 & 40 & 66 & \verb+OT2_lkrist01_2+ \\
L1489		& 1342249643 & 60 & 50 & \verb+OT2_lkrist01_2+ \\
L1551-IRS5	& 1342249648 & 40 & 68 & \verb+OT2_lkrist01_2+ \\
TMR1		& 1342249649 & 60 & 52 & \verb+OT2_lkrist01_2+ \\
HH46		& 1342210792 & 71 & 46 & \verb+GT1_abenz_1+ \\
DK Cha		& 1342201659 & 61 & 53 & \verb+KPGT_evandish_1+ \\
GSS30-IRS1	& 1342251067 & 31 & 73 & \verb+OT2_lkrist01_2+ \\
Elias29		& 1342251071 & 20 & 110 & \verb+OT2_lkrist01_2+ \\
Oph-IRS44	& 1342251068 & 31 & 79 & \verb+OT2_lkrist01_2+ \\
RCrA-IRS5A	& 1342253826 & 31 & 68 & \verb+OT2_lkrist01_2+ \\
RCrA-IRS7C	& 1342244583 & 15 & 180 & \verb+OT2_lkrist01_2+ \\
RCrA-IRS7B	& 1342244584 & 15 & 120 & \verb+OT2_lkrist01_2+ \\ \hline
\end{tabular}
\tablefoot{Integration times are on+off. The rms is measured in 1~km\,s$^{-1}$ channels.}
\end{center}
\end{table}

\begin{table*}
\caption{Gaussian fits to each line profile. \label{tab:gauss}}
\tiny
\begin{center}
\begin{tabular}{l c c c c c c c c c c c}\hline \hline
& & & \multicolumn{4}{c}{Free fit\tablefootmark{c}} && \multicolumn{4}{c}{Mottram fit\tablefootmark{d}} \\
\cline{4-7} \cline{9-12} \\
Source & $\varv_{\rm source}$\tablefootmark{a} & Comp.\tablefootmark{b} & $\varv_{\rm peak}$ & $FWHM$ & $T^{\rm peak}_{\rm MB}$ & $\int T_{\rm MB}$d$\varv$ && $\varv_{\rm peak}$ & $FWHM$ & $T^{\rm peak}_{\rm MB}$ & $\int T_{\rm MB}$d$\varv$ \\
& (km\,s$^{-1}$) && (km\,s$^{-1}$) & (km\,s$^{-1}$) & (K) & (K km\,s$^{-1}$) && (km\,s$^{-1}$) & (km\,s$^{-1}$) & (K) & (K km\,s$^{-1}$) \\ \hline
L1448-MM	& \phs5.2 	& B & \phs16.0 		& 54.8 	& 0.16 & 9.24 && \phs16.8 	& 44.6 	& 0.18 & 8.73 \\
			&		& O & $-$43.4 		& 30.5 	& 0.23 & 7.60 && $-$41.8 		& 39.8 	& 0.26 & 11.2 \\
			&		& O & \phs58.8 	& 15.4 	& 0.31 & 5.12 && \phs59.6 	& 23.0 	& 0.25 & 6.06 \\
N1333-IRAS2A	& \phs7.7	& B & \phs\phn8.1	& \phn6.8 	& 0.44 & 3.18 && \phs\phn8.0 	& \phn4.8 	& 0.36 & 1.83 \\
			& 		& O & $-$16.8	 	& 28.0 	& 0.10 & 2.92 && \phn$-$5.0 	& 39.9	& 0.09 & 3.97 \\
			& 		& O & \phs15.2	 	& 18.4	& 0.13 & 2.61 && \phs11.5 	& 13.7	& 0.17 & 2.47 \\
N1333-IRAS4A	& \phs7.2	& O & \phs\phn0.3 	& 11.6 	& 0.31 & 3.77 && \phn$-$0.8 	& \phn9.9 & 0.48 & 5.05 \\
			& 		& B/O & \phs\phn4.1 & 31.8 	& 0.43 & 14.7 && \phn\phs8.4 	& 18.0 	& 0.22 & 4.28 \\
			& 		& B &  			&  		&  	&  		&& \phs\phn9.9	& 41.4 	& 0.21 & 9.06 \\
N1333-IRAS4B	& \phs7.4	& O & \phs\phn6.2 	& \phn4.1 	& 0.72 & 3.11 && \phs\phn\ldots & \phn\ldots & \ldots & \ldots \\
			& 		& B & \phn\phs7.7 	& 19.5 	& 1.05 & 21.9 && \phs\phn8.0 	& 24.8 	& 0.78 & 20.6 \\
BHR71		& $-$4.4	& O & $-$67.8	 	& 17.9	& 0.19 & 3.64 && $-$57.4 		& 59.0 	&  &  \\
			& 		& O & $-$27.1 		& 37.0	& 0.23 & 9.26 &&  	&  &  &  \\
			& 		& B & \phn$-$6.3 	& \phn6.4 	& 0.82 & 5.56 && \phn$-$8.4 	& \phn6.4 &  &  \\
IRAS15398	& \phs5.1	& B & \phn\phs5.4 	& 10.1 	& 0.43 & 4.65 && \phn$-$0.4 	& 16.4 	&  &  \\
VLA1623		& \phs2.8	& O & \phs\phn0.9 	& \phn7.1 	& 0.73 & 5.55 &&  	&  &  &  \\
			& 		& B & \phs\phn2.2 	& 28.1 	& 0.10 & 2.99 &&  	&  &  &  \\
L483			& \phs5.2	& B & \phs\phn2.8 	& 12.2 	& 0.27 & 3.50 && \phn\phs3.2 	& 18.5 &  &  \\
			& 		& N & \phn\phs4.3 	& \phn3.2 	& 0.34 & 1.16 &&  	&  &  &  \\
Ser-SMM1	& \phs8.5	& O &  \phs10.2	& 10.6 	& 0.82 & 9.27 &&  	&  &  &  \\
			& 		& O & \phn\phs5.9 	& \phn7.9 	& 2.95 & 24.9 &&  	&  &  &  \\
			& 		& B & \phs\phn6.1 	& 19.8 	& 1.91 & 40.3 &&  	&  &  &  \\
Ser-SMM4\tablefootmark{e}	& \phs8.0	& ? &  	&  &  &  &&  	&  &  &  \\
Ser-SMM3	& \phs7.6	& O & \phn\phs1.1 	& 15.6 	& 0.22 & 3.64 &&  	&  &  &  \\
			& 		& O & \phs16.0 	& 13.9 	& 0.27 & 3.93 &&  	&  &  &  \\
B335			& \phs8.4	& B & \phn\phs6.5 	& 31.8 	& 0.12 & 4.00 && \phn\phs9.9 	& \phn4.2 & 0.12 & 0.52 \\
			& 		&  &  &  &  &  && \phn\phs7.9 	& 40.9 & 0.09 & 3.75 \\
L1157		& \phs2.6	& O & $-$13.2 		& 55.5 	& 0.11 & 6.65 &&  	&  &  &  \\
			& 		& B & \phn\phn1.1 	& \phn5.9 	& 0.17 & 1.10 &&  	&  &  &  \\
L1489		& \phs7.2	& B & \phn\phs6.1 	& \phn9.8 	& 0.41 & 4.25 && \phn\phs3.7 	& 20.0 &  &  \\
L1551-IRS5	& \phs6.2	& B & \phn\phs3.4 	& 14.9 	& 0.36 & 5.76 &&  	&  &  &  \\
			& 		& N & \phs\phn6.1 	& \phn3.6 	& 0.86 & 3.13 && \phn\phs6.7 	& \phn4.3 &  &  \\
TMR1		& \phs6.3	& B & \phn\phs4.8 	& \phn5.5 	& 1.00 & 5.90 &&  	&  &  &  \\
HH46		& \phs5.2	& B & \phn\phs4.6 	& \phn6.5 	& 0.16 & 1.08 && \phs10.5 	& 23.5 &  &  \\
DK Cha		& \phs3.1	& O & --15.5 		& 31.7 	& 0.13 & 4.24 &&  	&  &  &  \\
			& 		& N & \phn\phs2.3 	& \phn4.8 	& 0.76 & 3.87 && \phn\phs3.5 	& \phn4.5 	& 0.60 & 2.87 \\
GSS30-IRS1	& \phs3.5	& B & \phn\phs0.8 	& 11.4 	& 0.89 & 10.8 && \phn\phs2.2 	& 14.5 	& 1.04 & 16.1 \\
			& 		& N & \phn\phs1.8 	& \phn4.0 	& 1.51 & 6.47 && \phs\phn3.0 	& \phs2.6 	& 1.17 & 3.24 \\
Elias29		& \phs4.3	& N & \phn\phs4.5 	& \phn4.5 	& 1.16 & 5.55 && \phn\phs5.2 	& 13.5 	& 1.98 & 28.4 \\
			& 		& B & \phn\phs4.1 	& \phn9.6 	& 1.80 & 18.4 &&  	&  &  &  \\
Oph-IRS44	& \phs3.5	& N & \phn\phs3.5 	& \phn5.3 	& 1.14 & 6.39 &&  	&  &  &  \\
			& 		& B & \phn\phs3.5 	& 13.8 	& 1.36 & 19.9 &&  	&  &  &  \\
RCrA-IRS5A	& \phs5.7	& B & \phn\phs4.9 	& \phn7.8 	& 0.70 & 5.81 &&  	&  &  &  \\
RCrA-IRS7C	& \phs5.5	& N & \phn\phs4.8 	& \phn5.4 	& 4.46 & 25.4 &&  	&  &  &  \\
			& 		& B & \phn\phs5.0 	& 11.1 	& 2.83 & 33.4 &&  	&  &  &  \\
RCrA-IRS7B	& \phs5.9	& N & \phn\phs5.7 	& \phn4.7	& 2.55 & 12.9 &&  	&  &  &  \\
			& 		& B & \phn\phs4.7 	& 10.3 	& 0.87 & 9.49 &&  	&  &  &  \\ \hline
\end{tabular}
\tablefoot{
\tablefoottext{a}{Source velocity, from \citet{yildiz13}.}
\tablefoottext{b}{Component identifier: B is for the broad cavity shock component, O for the offset spot shock component, and N for the narrow shock component; see Table \ref{tab:component} for details.}
\tablefoottext{c}{Gaussian fit where all parameters are left free.}
\tablefoottext{d}{Gaussian fit where the values of $\varv_{\rm peak}$ and $FWHM$ are from \citet{mottram14}.}
\tablefoottext{e}{This profile is not decomposed into Gaussians because the profile is triangular.}
}
\end{center}
\end{table*}

\end{document}